\documentclass[journal,twocolumn]{IEEEtran}

\usepackage{xcolor}

\usepackage{mathtools}
\usepackage{epsfig,makeidx,color}
\usepackage{amsmath,amssymb,bbm}
\usepackage{cite,graphicx,lipsum}
\usepackage{enumerate}
\usepackage[switch,pagewise]{lineno}
\usepackage{hyperref}
\hypersetup{
        colorlinks = true,
        citecolor=blue,
}

\usepackage{caption}
\usepackage{subcaption}

\usepackage{listings}
\usepackage[most]{tcolorbox}

\def\cX{{\cal X}}

\def\cQ{{\cal Q}}

\def\uE{{\mathbb E}}

\DeclareMathOperator*{\argmin}{\arg\!\min}

\newtheorem{myexample}{\it Example} 

\def\be{ \begin{equation} }
\def\ee{ \end{equation} }
\def\bea{ \begin{eqnarray} }
\def\eea{ \end{eqnarray} }

\def\bx{{\bf x}}

\def\b0{{\bf 0}}

\def\cP{{\cal P}}
\def\cQ{{\cal Q}}

\def\cI{{\cal I}}

\def\sH{{\sf H}}
\def\sI{{\sf I}}

\ifCLASSOPTIONonecolumn
  \newcommand{\figwidth}{0.50\columnwidth}
  
\else
  \newcommand{\figwidth}{0.90\columnwidth}
  
\fi

\begin{document}

\title{A Unified Approach to Semantic Information and Communication based on Probabilistic Logic}

\author{Jinho Choi, Seng W. Loke, and Jihong Park\\
\thanks{The authors are with
the School of Information Technology,
Deakin University, Geelong, VIC 3220, Australia
(e-mail: \{jinho.choi,seng.loke,jihong.park\}@deakin.edu.au).
This research was supported
by the Australian Government through the Australian Research
Council's Discovery Projects funding scheme (DP200100391).}}


\maketitle
\begin{abstract}
Traditionally, studies on technical communication (TC) are based on stochastic modeling and manipulation. This is not sufficient for semantic communication (SC) where semantic elements are logically connected, rather than stochastically correlated. To fill this void, by leveraging a logical programming language called probabilistic logic (ProbLog), we propose a unified approach to semantic information and communication through the interplay between TC and SC. 
On top of the well-established existing TC layer, we introduce a SC  layer that 
utilizes knowledge bases of communicating parties for the exchange of semantic information. These knowledge bases are logically described, manipulated, and exploited using ProbLog. To make these SC and TC layers interact, we propose various measures based on the entropy of a clause in a knowledge base. These measures allow us to delineate various technical issues on SC such as a message selection problem for improving the knowledge base at a receiver. Extending this, we showcase selected examples in which SC and TC layers interact with each other while taking into account constraints on physical channels and efficiently utilizing channel resources.
\end{abstract}


\ifCLASSOPTIONonecolumn
\baselineskip 28pt
\fi

\section{Introduction}

It may not be an exaggeration to say that the successful development of various communication systems, including fifth generation (5G) cellular systems, has been based on Shannon's theory, which is called information theory \cite{Shannon}.
Information theory has influenced the development of communication systems as well as various other fields (e.g., statistics, biology, and so on). In his theory, information is characterized as randomness in variables. This allows one to calculate the fundamental limits and performance of communication, and to design efficient compression and transmission schemes through noisy channels. Despite the success in this domain of \emph{technical communication (TC)}, since its introduction in 1948, Shannon theory's ignorance about the meanings of information \cite{Weaver53} has long been tackled particularly in the field of the philosophy of information. Meanwhile, overcoming this limitation of Shannon theory has recently been regarded as one of the key enablers for the upcoming sixth generation (6G) communication systems \cite{Strinati21,dde,seo2021semantics}. 

To fill this void, it requires to develop a theory on meaningful information, i.e., \emph{semantic information}, as well as a novel communication technology based on semantic information, i.e., \emph{semantic communication (SC)}. For SC, existing works can be categorized into model-free methods leveraging machine learning \cite{dde}, and model-based approaches that quantify semantic information \cite{Bao11} or specify the emergence of meanings through communication \cite{seo2021semantics}. Our work falls into the latter category in the hope of unifying our analysis on SC with the existing model-based analysis on TC.

In regard to semantic information, there are two different views in the philosophy of information. One angle focuses on measuring semantic similarity \cite{Floridi05,Floridi08}, which often encourages an entirely new way to define meaningful information. For instance, each meaning can be identified as a group that is invariant to various nuisances or a category \cite{belfiore2021topos}, across which semantic similarity can be compared.
The other end of the spectrum focuses on quantifying semantic uncertainty \cite{Adriaans10}, in a similar way to Shannon theory where message occurrences are counted to measure semantic-agnostic uncertainty. As an example, Shannon information can be extended to semantic information by leveraging the theory of inductive probability \cite{Carnap50} (see also \cite{Hailperin84,Williamson02}). This allows to measure the likelihood of a sentence/clause's truth using logical probability, upon which an SC system can be constructed~\cite{Bao11}. Our view is aligned with the latter angle (i.e., like \cite{Bao11}, a probabilistic logic approach is taken), 
while we focus on making SC interact with TC under Shannon theory.





In particular, in this paper, 
we consider an approach to SC based on the theory of probabilistic logic assigning probabilities to logical clauses \cite{Carnap50,Nilsson86}. This allows to make inferences over clauses and to quantify their truthfulness or provability in a probabilistic way. We showcase that the process of inference and its provability analysis can be performed using the \emph{probabilistic logic programming language (ProbLog)}\footnote{ProbLog tools are available in: \url{https://dtai.cs.kuleuven.be/problog}.}, a practical logic-based probabilistic programming language that has been widely used in the field of symbolic artificial intelligence (AI). 

Furthermore, based on  \cite{Bar-Hillel53,Adriaans10}, we consider a two-layer SC system comprising: (i) the conventional TC layer where data symbols can be transmitted without taking into account their meanings; and (ii) an SC layer where one exploits semantic information that can be obtained from a background knowledge or by updating a knowledge base. We demonstrate the interaction between TC and SC layers with selected examples showing how SC improves the efficiency of TC, i.e., \emph{SC for TC}, as well as how to design TC to achieve maximal gains in SC under limited communication resources, i.e., \emph{TC for SC}. 
For simplicity and consistency throughout the paper, we confine ourselves to a simple scenario where a human user or an intelligent device stores logical clauses in a knowledge base and intends to improve the knowledge by seeking answers to a number of queries.



The main contributions of the paper are as follows.
\begin{enumerate}
    \item Based on probabilistic logic, we characterize knowledge bases for semantic information and define various entropy-based measures, which allow us to model semantic compression and security.
\item For a SC system consisting of SC and TC layers, we address various issues through interactions between SC and TC subject to constraints of physical channels including a message selection problem. Few numerical examples are studied to illustrate the proposed approaches.
\item Open issues and challenges are identified for further research in the future.
\end{enumerate}
Note that this paper is an extended version of \cite{CLP_22}.

The paper is organized as follows. Once we provide a background in Section~\ref{S:Background}, in Section~\ref{S:EKB}, we present various aspects of semantic information and knowledeg bases based on probabilistic logic and introduce key measures.  With the developed measures, in Section~\ref{S:SCTC}, we address key issues to build a SC system consisting of TC and SC layers by explaining how TC and SC layers interact subject to various constraints of physical channels. Numerical results on two exemplary SC use cases are presented in Section~\ref{S:NR}, and open issues and challenges are discussed in Section~\ref{S:OIC}. We conclude the paper with a few remarks in Section~\ref{S:Conc}.

\section{Backgrounds}    \label{S:Background}

In this section, we present a background on information theory \cite{CoverBook} and probabilistic logic \cite{Nilsson86} \cite{Williamson02}.

\subsection{Classical and Semantic Information Theory}

Although information theory originally started as a mathematical theory for communications, it
has been applied in diverse fields ranging from biology to neuroscience.
In information theory, random variables are used to represent symbols to be transmitted. The entropy of a random variable, denoted by $X$, is the number of bits required to represent it, which is given by $\sH(X) = - \sum_x \Pr(x) \log \Pr(x) = \uE[ - \log \Pr(X)]$ (taking $\log$ to base 2 in the rest of the paper) when $X$ is a discrete random variables, where $\Pr(x)$ stands for the probability that $X = x$ and $\uE[\cdot]$ represents the statistical expectation. 
The entropy of $X$ can also be interpreted as the amount of information of $X$.

The joint entropy of $X$ and $Y$ is defined as $\sH(X, Y) = \uE[-\log \Pr(X,Y)]$ and the conditional entropy is given by
$$
\sH(X\,|\, Y) = \uE[ - \log \Pr(X\,|\, Y)] = \sH(X,Y) - \sH(Y).
$$
The mutual information between $X$ and $Y$ is defined as $\sI(X;Y) =
\uE\left[\log \frac{\Pr(X,Y)}{\Pr(X) \Pr(Y)} \right]$. It can also be shown that $\sI(X;Y) = \sI(Y;X) = \sH(X) - \sH(X\,|\, Y) = \sH(Y) - \sH(Y\,|\, X)$. 
If $X$ and $Y$ are assumed to be the transmitted and received signals over a noisy channel, respectively, $\sI(X;Y)$ can be seen as the number of bits that can be reliably transmitted over this channel. Thus, $\max_{\Pr(x)} \sI(X;Y)$ is called the channel capacity that is the maximum achievable transmission rate for a given channel that is characterized by the transition probability $\Pr(Y\,|\, X)$. 

As pointed out in \cite{Bar-Hillel53}, information theory is not interested in the content or meaning of the symbols, but quantifying the amount of information based on the frequency of their occurrence (i.e., the distribution of symbols as random variables). For example, $\sH(X)$ is to measure the uncertainty of information or number of bits to represent a symbol $X$ regardless of what $X$ means. 
However, this does not mean that information theory is useless in dealing with the meaning or content of information as will be discussed in the paper.

To fill this void, it requires to develop a theory on meaningful information, i.e., semantic information. In regard to semantic information, there are two different views in the philosophy of information. One angle focuses on measuring semantic similarity \cite{Floridi05,Floridi08}, which often encourages an entirely new way to define meaningful information. For instance, each meaning can be identified as a group that is invariant to various nuisances (e.g., a so-called topos in category theory \cite{belfiore2021topos}), across which semantic similarity can be compared.
The other end of the spectrum focuses on quantifying semantic uncertainty \cite{Adriaans10}, in a similar way to Shannon theory where message occurrences are counted to measure semantic-agnostic uncertainty. As an example, Shannon information can be extended to semantic information by leveraging the theory of inductive probability \cite{Carnap50} (see also \cite{Hailperin84,Williamson02}). This allows to measure the likelihood of a sentence/clause's truth using logical probability, upon which an SC system can be constructed~\cite{Bao11}. Our view is aligned with the latter angle (i.e., like \cite{Bao11}, a probabilistic logic approach is taken), 
while we focus on making SC interact with TC under Shannon theory.

\subsection{Deterministic and Probabilistic Logic} 

Reasoning about the truth of a sentence is the simplest type of logic, called propositional logic.  Treating this as the zero-th order logic, the first-order logic can describe ordinary logic by parsing out and dividing each sentence into meaningful clauses \cite{Aho94}. In the first-order logic, each clause consists of constant symbols (e.g., alphabets), logical operators (e.g., Boolean algebra such as AND $\wedge$, OR $\lor$, and NOT $\neg$), and non-logical predicates (e.g., $x$ ``is the father of" $y$). Programming in Logics (Prolog) aims to describe the first-order logic using a programming language, which has been widely used for computational linguistics and symbolic artificial intelligence (AI) such as IBM Watson \cite{Clocksin03}. In Prolog, each clause is in a form of {\tt Head~:-~Body} which is read as ``Head is True if Body is True." However, Prolog can only describe deterministic logic although the world is full of uncertainty. To overcome this limitation, thanks to the notion of probabilistic logic \cite{Nilsson86} \cite{Williamson02},
ProbLog introduces the notion of a probability $p$ to each clause that is now in a form of {\tt p::Head~:-~Body}. This probability $p$ can be, for instance, annotated by a programmer, which indicates the programmer's degree of belief in the clause.

In this paper, we focus on exchanging logical clauses and making probabilistic inferences based on the clauses written in ProbLog. In particular, for facts $a$ and $b$, where $a$ is assigned probability $p_a$ and $b$ is assigned probability $p_b$, we have probability of $a \wedge b$ computed as the product of the probabilities, i.e. $p_a \cdot p_b$, and $a \lor b$ computed as $1-(1-p_a)\cdot(1-p_b)$ since $a \lor b = \neg (\neg a \wedge \neg b)$. 
Similar calculations can be applied with deductive reasoning, e.g., suppose  we have the rule $r$ of the form $a \rightarrow b$ (where ``$\rightarrow$'' is ``implies'')  annotated with probability $p_r$ and $a$ with probability $p_a$, then we can infer $b$ with probability $p_r \cdot p_a$. In ProbLog, a clause $a \rightarrow b$ with probability $p$ is written as {\tt p::b~:-~a}, where ``{\tt :-}'' is read as ``if''.


In general,  a knowledge base $K$ is regarded as a set of clauses (where a clause is a rule or a fact). Given a rule of the form $a \rightarrow b$, the head of the rule is $a$ and the body is $b$. Note that a fact is basically a rule of the form $a \rightarrow true$, which can just be written as $a$. 
One can make inferences about the truth value of a query $q$, provided that $q$ matches the head of a clause in $K$ with the outcome being  the probability of $q$. If $q$ does not match any head of a clause in $K$, $K$ cannot say anything about $q$. We denote the probability of $q$ computed as the answer when posed as a query to the knowledge base $K$ by $p[K \vdash q]$. We assume that inferences made will be as defined by the semantics of ProbLog.

In addition, for the purposes of the discussion in this paper, we consider mostly the propositional logic fragment of ProbLog for simplicity (and if variables are involved in some examples, we assume that their values range over a finite set, i.e., they are just abbreviations for a finite set of propositional clauses, so that the set of queries that can be answered via a knowledge base is finite).

\section{Entropy and Knowledge Bases: Communicating  Informative Messages} \label{S:EKB}

In this section, we discuss various  aspects of semantic information
(e.g., semantic compression and security) after  quantifying the uncertainty of knowledge bases using the entropy of a clause.

\subsection{Entropy of a Clause}
We  consider the entropy $\sH_f$ of a given clause  $c$  whose truth value can be considered as a random variable with outcomes ``true'' with probability $p_c$, and ``false'' with probability $1-p_c$,   as follows:
\[
 \sH_f(c) = - \left(p_c  \log (p_c) + (1-p_c)  \log(1-p_c) \right).
\]
Here, the subscript $f$ is used to differentiate the entropy of a random variable from that of a clause.

When a given query $q$ is posed to the knowledge base $K$, and suppose a probability $p_q$ is computed with respect to   $K$, i.e., when $q$ matches a head of a clause in $K$, as in the semantics of ProbLog, then  $p_q = p[K \vdash q]$, and we denote the entropy of $q$ with respect to $K$ as   $\sH_f^K(q)$, i.e.:
\[
 \sH_f^K(q) = - \left(p_q  \log (p_q) + (1-p_q)  \log(1-p_q) \right).
\]
Note that if $q$ does not match the head of any clause in $K$, then the result of the query is undefined; alternatively, for an application, this can be set to $0.5$ (i.e., a random guess). 


\subsection{Uncertainty of a Knowledge Base}

Let $\mathcal{H}_K$ denote the set of the terms which are the heads of all clauses in $K$. We consider the heads of the clauses as these would correspond to the set of different queries that the knowledge base can compute a meaningful probability for.

Given a knowledge base $K$, we can then define an uncertainty measure $U_{KB}$ of $K$ as follows (which takes into  account the entropy of answers it computes, i.e.,  the average  entropy of queries computable from $K$):
\be \label{Eq:AvgEntropy}
 U_{KB}(K) = \frac{1}{|\mathcal{H}_K|} \sum_{q \in \mathcal{H}_K} \sH_f^K(q).
\ee 
Ideally, if a knowledge base can answer all its queries with certainty (probability 1, i.e., true with probability 1 or false with probability 1), then $ U_{KB}(K)=0$ (assuming that $0\cdot \log(1/0)=0$), while it is $1$ in the worse case.
\begin{tcolorbox}[enhanced,sharp corners,width={\columnwidth},breakable,colback=white, boxsep=1pt,left=2pt,right=2pt,top=3pt,bottom=3pt]
\begin{myexample}
Suppose we have a knowledge base $K$ as follows, in ProbLog:
\begin{verbatim}
0.2::a.
0.3::b.
0.5::a :- b.
\end{verbatim}
The set of the heads of all clauses in $K$ is $\{a,b\}$;  the possible queries $K$ can answer are $a$ and $b$, i.e.
$p[K \vdash a]= 1-(1-0.2)(1-(0.3)(0.5)) = 0.32$, and 
$p[K \vdash b]=0.3$. Thus,
\begin{align*}
 U_{KB}(K)  = \frac{\sH_f^K(a) + \sH_f^K(b)}{2} 
= \frac{0.904 + 0.881}{2} = 0.8925.
\end{align*}
\end{myexample}
\end{tcolorbox}

\subsection{Sender's Message Choice Problem} \label{Sec:SenderChoice}

Consider a network or a multiuser system consisting of multiple users.
Each user may wish to improve their knowledge bases and 
communication\footnote{In this section, we assume that TC is ideal. In Section~\ref{S:SCTC}, we will consider how SC and TC interact.} 
plays a crucial role in reducing the uncertainty of a knowledge base. To illustrate this, suppose that Alice has a set $L$ of clauses and Bob has a knowledge base $K$. In order to minimize the average entropy of $K$, Alice can choose and send a message $m\in L$ to Bob, i.e.,
\be
m = \argmin_{l \in L} U_{KB}(K\cup\{l\}). \label{Eq:SenderChoice}
\ee

However, this requires Alice to have complete knowledge of~$K$. Alternatively, Alice might have a statistical approximation $A_i$ of $K$ in which $A_i \approx K$ with probability $p_{A_i}$. In this case, Alice's choice of $m$ is recast as:
\be
 m = \argmin_{l \in L} \sum_{i} p_{A_i} U_{KB}(A_i\cup\{l\}).
\ee
To realize this idea, one way is to allow Bob to keep feeding the entropy of $K$ back to Alice. Then, throughout iterative communication, Alice can gradually improve the accuracy of~$A_i$.







\subsection{Receiver's Message Assimilation Problem}

In parallel with Alice's choice of communication message $m$ as discussed in Section~\ref{Sec:SenderChoice}, Bob is also able to reduce the uncertainty of the knowledge base $K$ by adjusting the updating rule of $K$ upon receiving $m$, i.e., assimilation of $m$. In \eqref{Eq:SenderChoice}, the assimilation is given by simply adding the received message to $K$, i.e., $K\cup\{m\}$. Generalizing this, Bob's message assimilation problem is cast as: $\min_{\circ \in \mathcal{A} } U_{KB}(K \circ\{m\}) $, 
where $\circ$ identifies an operator among a set $\mathcal{A}$ of assimilation operators.

The aforementioned simple addition can be an assimilation operator, i.e., $\cup \in \mathcal{A}$. Additionally, we introduce an assimilation operator maximizing the freshness of each clause in a way that: on receiving a new message (or clause) $m'$ of the form \mbox{\tt $p_{m'}$::$l$}, if $K$ includes clauses  $m$ (of the form \mbox{\tt $p_m$::$l$}) differing from $m'$ in only the associated  probability $p_m$, it replaces all such clauses of $m$ with the newly received $m'$, resulting in the updated knowledge base $K'=K \backslash \{\mbox{\tt $p_m$::$l$}\} \cup \{\mbox{\tt $p_{m'}$::$l$}\}$ with replacement; otherwise, it follows the simple addition rule. To describe this, we define an assimilation operator $\odot$ that satisfies:
\begin{align*}
K \odot \{\mbox{\tt $p_{m'}$::$l$}\}  = 
\begin{dcases}
    K' ,& \text{if }  \mbox{\tt $p_m$::$l$} \in K,\\ &  \text{for some $p_m$} \\
     K  \cup \{\mbox{\tt $p_{m'}$::$l$}\},              & \text{otherwise.}
\end{dcases}
\end{align*}

Furthermore, we introduce another assimilation rule that aims to minimize the entropy of each query to be asked to $K$. To this end, $K$ remains unchanged if the received $m'$ doesn't help decrease the entropy for the query corresponding to the head $h_{m'}$ of $m'$, where the clause $m'$ is in the form of \mbox{\tt $p_{m'}$::$h_{m'}$:-$b_{m'}$}, i.e. $l = \mbox{\tt $h_{m'}$:-$b_{m'}$}$. This rule is described using an assimilation operator $\oplus$ that is defined as:
\begin{align*}
    K \oplus \{\mbox{\tt $p_{m'}$:: $l$} \} = 
\begin{dcases}
    K', &   \text{if } \mbox{\tt $p_m$::$l$} \in K,\\ & \text{for some $p_m$}, \\  
        & ~\&~ \sH_f^{K'}\!(h_{m'}) < \sH_f^K\!(h_{m'}) \\
      K, & \text{if } \mbox{\tt $p_m$::$l$} \in K,\\ & \text{for some $p_m$}, \\  &  ~\&~ \sH_f^{K'}\!(h_{m'}) \geq \sH_f^K\!(h_{m'}) \\
     K ~~\cup & \{\mbox{\tt $p_{m'}$::$l$} \}, \text{otherwise.} 
\end{dcases}
\end{align*}
Given the assimilation operator $\cup$, $\odot$, or $\oplus$, the resultant changes in the average entropy of $K$ will be elaborated on  in Section~\ref{Sec:EntropyChange}.  Furthermore, for simplicity, 
$\cup$ will be used to represent the assimilation operators discussed above (i.e., $\odot$, or $\oplus$).

\subsection{Semantic Content of a Message} \label{Sec:EntropyChange}
We can define the notion of the {\em semantic content} $\mathcal{S}$ of  a message (where a message in this case is a clause labelled with a probability) with respect to the receiver's background knowledge base $K$ as follows (as the change in average entropy of a knowledge base with respect to its queries):
\be 
\mathcal{S}_K(m) =  U_{KB}(K\cup\{m\}) - U_{KB}(K)  .
    \label{EQ:S_K} 
\ee 


Each message changes $U_{KB}$ and the receiver wants 
to decrease the entropy, i.e., $\mathcal{S}_K(m) \leq 0$, or wants $\mathcal{S}_K(m)$ to be as low as possible, as the message $m$ should  decrease the average entropy in computed queries (of course, it could also increase the average entropy!).
In the following example, we show why this definition helps.
\begin{tcolorbox}[enhanced,sharp corners,width={\columnwidth},breakable,colback=white, boxsep=1pt,left=2pt,right=2pt,top=3pt,bottom=3pt]
\begin{myexample} 
Suppose Alice has  a knowledge base $K$ as follows, in ProbLog:
\begin{verbatim}
0.3::b.
0.5::a :- b.
\end{verbatim} 
Suppose Alice receive the labelled clause {\tt 0.2::m}, i.e., {\tt m} labelled with probability $0.2$ forming $K'$ as follows:
\begin{verbatim}
0.3::b.
0.5::a :- b.
0.2::m.
\end{verbatim} 
Then, $p[K' \vdash a]=0.15$, $p[K' \vdash b]=0.3$, and $p[K' \vdash m]=0.2$ and so $U_{KB}(K') = \frac{1}{3} (0.60984 + 0.88129 + 0.721928) = 0.738$. We have:
\begin{align*}
\mathcal{S}_K(m)  & =   U_{KB}(K\cup\{0.2::m\}) - U_{KB}(K)  \\
  & = 0.738 - 0.746 \approx -0.008.
\end{align*}
The uncertainty in the knowledge base with respect to the queries it can answer has decreased - which is what we expect when Alice receives a clause with a lower entropy relative to the existing  clauses in $K$. 
Also,  if instead Alice received {\tt 0.9::b}, then  Alice's knowledge base becomes:
\begin{verbatim}
0.9::b.
0.5::a :- b.
\end{verbatim} 
And $p[K' \vdash a]=0.45$, $p[K' \vdash b]=0.9$, that is, we have:
\begin{align*}
\mathcal{S}_K(m)  & = U_{KB}(K\cup \{0.9::b\}) - U_{KB}(K) \\
  & = 0.731 - 0.746 \approx -0.015 .
\end{align*}
The uncertainty in the knowledge base with respect to the queries it can answer has decreased - which is what we expect when Alice receives a clause with a lower entropy replacing an  existing  clause in $K$.
By assimilating {\tt 0.9::b}, we can have:
\begin{verbatim}
0.9::b.
0.3::b.
0.5::a :- b.
\end{verbatim} 
where  $p[K' \vdash a]=0.465$, $p[K' \vdash b]=0.93$, and
\begin{align*}
\mathcal{S}_K(m)  & = U_{KB}(K\cup\{0.9::b\}) - U_{KB}(K) \\
  & = 0.0.681 - 0.746 \approx -0.065 
\end{align*}
which also shows a decrease in average entropy.
\end{myexample} 
\end{tcolorbox}

\subsection{Inference Can Reduce the Need for Communication} 
    \label{SS:IRNC}

Suppose there is no background knowledge, i.e., $K =\emptyset$.
Then, the uncertainty of a query becomes $\sH^{\emptyset}_f(q) = 1$, i.e. the truth or falsity of $q$ is merely a random guess. But with a knowledge base $K\not= \emptyset$, we expect to have:
$\sH_f^K(q) \leq \sH^{\emptyset}_f(q)$.
Furthermore, for two different knowledge bases, $K$ and $K^\prime$,~if
$$
\sH_f^{K} (q) \le \sH_f^{K^\prime} (q),
$$
then we say $K$ is less uncertain than $K^{\prime}$ with respect to query $q$. For the case that $K^\prime \subseteq K$, we can easily show that 
$\sH_f^{K} (q) \le \sH_f^{K^\prime} (q)$.
 
This can lead to a reduction in the need to obtain information about $q$ given that we can make inferences about $q$ with $K$. For example, suppose $\sH_f^K(q) > 1 - \delta$, where $\delta > 0$, is good enough, then there is no need to receive further information about $q$. In fact, with respect to $q$, we want only to receive information to reduce the entropy for $q$, that is, we want only to receive message $m \notin K$ such that:
\[
\sH^{K \cup \{m\}}_f(q) \leq \sH^{K}_f(q) .
\]
This can also be generalized if there is a set of available messages, say $U$, as follows:
\be 
m^* = \argmin_{m \in U} \sH^{K \cup \{m\}}_f(q).
    \label{EQ:mmin}
\ee 
Here, $m^*$ is the best message among those in $U$ to reduce the entropy for $q$. This implies that one might want to consider the consequences  of receiving and assimilating a message (or from the sender side, the implications of sending a message) on the uncertainty of a knowledge base (whether it would increase or decrease the entropy with respect to $q$ or with respect to the overall uncertainty of a knowledge base as defined above). We illustrate this idea further later in the paper.


\subsection{Communicating a Knowledge Base Efficiently: a Notion of Semantic Compression}

If the sender has an entire knowledge base to send, then the sender can achieve possible compression by sending the minimum number of clauses (assuming a standard fixed number of bits to send a clause) equivalent to the query-answering capability of the knowledge base. 

We used $\mathcal{H}_K$,  the heads of all clauses in $K$, crudely to represent the set of queries answerable by a knowledge base but more general measures can be defined based on what can be inferred from a knowledge base (e.g., the immediate consequence operator~\cite{10.1145/183432.183528}). 

Let $\mathcal{P}(K)$ denote the  set of queries answerable using knowledge base $K$, then two knowledge bases $K$ and $K'$ are equivalent  provided they can answer exactly the same queries:
$\mathcal{P}(K) = \mathcal{P}(K')$,
and for each $q \in \mathcal{P}(K)$, both the knowledge bases compute the same results $p[K \vdash q] = p[K' \vdash q]$. Denoting by $K_{eq}$ the set of all  knowledge bases equivalent to $K$, clearly, the sender should send $K_{min}$ to the receiver, which is given by
$$
 K_{min} = \argmin_{K' \in K_{eq}} |K'|,
$$
where $|K'|$ denotes the cardinality of $K'$, i.e., the number of clauses in $K'$. In practice, if this is hard to compute, the sender, wanting to send $K$, can try to perform {\em semantic compression} by finding a $K'$ such that  $K' \in K_{eq}$ and $|K'| < |K|$.

In fact, the conditions above can be weakened, if the sender has $K$ and the receiver has some tolerance, then, given some threshold of tolerances $\delta$ and $\epsilon$, suppose we have a knowledge base $K'$ such that 
$|\mathcal{P}(K)| - |\mathcal{P}(K')| = \delta |\mathcal{P}(K)|$ for a finite $|\mathcal{P}(K)| > 0$,
or
$\frac{|\mathcal{P}(K')| } { |\mathcal{P}(K)|} \ge 1 - \delta$,
and  that deviates from $K$ by computing potentially different though similar probabilities as $K$ for each query, that is, for each $q \in \mathcal{P}(K) \cap \mathcal{P}(K')$: 
$$
\big| p[K \vdash q] - p[K' \vdash q] \big| < \epsilon
$$ 
which also implies that, for some $\epsilon'$, restricted to commonly answerable queries, 
$\big|U_{KB}(K) - U_{KB}(K') \big|  < \epsilon'$.
Potentially, $K'$ can be a subset of $K$ by removing clauses, a ``compressed'' form for $K$.

We note that one can also define the {\em semantic content of a complex message} comprising, not just a single clause, but a set of clauses $M$ (i.e., where a set if clauses is a knowledge base), generalizing from (\ref{EQ:S_K}):
\be 
\mathcal{S}_K(M) =  U_{KB}(K\cup M) - U_{KB}(K)  .
    \label{EQ:S_KM} 
\ee 

We have seen that the sender who knows  the receiver has knowledge $K$  
can exploit this fact to reduce the amount of data that needs to be sent to the receiver, 
while communicating the same semantic content. In effect, one can compute the following, with respect to receiver knowledge $K$ and target semantic content  $T$ that the sender wants to communicate to the receiver:
$$
 M_{min} = \argmin_{M \in E_T} |M| ,
$$
where $E_T$ denotes the set of  complex messages having content $T$, i.e., $E_T = \{M\,|\, \mathcal{S}_K(M)=T\}$. This can be viewed as a form of semantic compression that is relative to the semantic content (as defined in  (\ref{EQ:S_KM})) to be communicated.

\subsection{Improved Security via Semantic Messages}

As we have seen, the semantic content of a message helps reduce the receiver's uncertainty about one or more queries.
We can then define a notion {\em semantically secure messages}, in that, without the receiver's knowledge base, someone who has gotten hold of the message might not be able to use it to answer a query (or a set of queries).

For example, suppose Eve has knowledge base $K_{\rm E}$ and Alice sends a message $m$ to Bob, who has knowledge base $K_{\rm B}$. With respect to a query $q$, we can represent the fact that Eve has little use for the message provided as follows:
\be
\sH_f^{K_{\rm E}}(q) = \sH_f^{K_{\rm E} \cup \{m\}}(q) .
    \label{sec:ignorance} 
\ee
In other words, suppose $\sH_f^{K_{\rm E}}(q) = 1$, and  Eve managed to intercept the communication and gain the message $m$ (and forwards it to Bob pretending that nothing has happened as a man-in-the-middle attack), but combined with her knowledge base $K_{\rm E}$, Eve is still just as uncertain about  $q$ as before.

However, Bob who receives $m$, who has $K_{\rm B}$, finds the message meaningful, that is, with respect to $q$:
\be
\sH_f^{K_{\rm B} \cup \{m\}}(q) <\sH_f^{K_{\rm B}}(q).
\label{sec:useful} 
\ee
Hence, as long as Bob and Alice have an a priori shared context, as represented by knowledge base $K_{\rm B}$ that Bob has and Alice knows that Bob has $K_{\rm B}$, then, it might be possible for Alice to transmit $m$ so that Eve (who does not know $K_{\rm B}$), an eavesdropper, will not be able to make much use of it, with respect to some ``sought  after'' answer for $q$.

Note that one can see this as analogous to the typical security encryption  scenario:  $q$ is the plaintext  encoded as the ciphertext $m$ using some key $k$, then Bob who has knowledge of the key $k$ can decrypt $m$ to know $q$, but Eve, after getting hold of $m$,  does not have $k$ and cannot use it obtain $q$. But there are key differences. There could be multiple ways to infer $q$ with different sets of clauses.
$K_{\rm B}$ and $K_{\rm E}$ may have different clauses but both could allow some inferences about $q$. 
Alice needs to ensure that $K_{\rm E}$ is such that \eqref{sec:ignorance} and $K_{\rm B}$ is such that  \eqref{sec:useful} before sending $m$.


We can consider semantic information security based on the previous discussion. Conventional information security \cite{Bloch11} \cite{Csiszar11} is based on different channel reliability (e.g., the eavesdropper channel is a degraded channel in wiretap channel models). 
On the other hand, semantic information security is based on the different reliability of knowledge bases.

Define the semantic mutual information between query $q$ and message $m$ with respect to knowledge base $K$ as
\be 
\sI_f^K [q;m] = \sH_f^K (q) - \sH_f^{K \cup \{m\}} (q).
\ee 
Like the mutual information, this semantic mutual information is non-negative, $\sI_f^K [q;m] \ge 0$, and upper-bounded by $\sH_f^K (q)$, i.e., 
$$
0 \le \sI_f^K [q;m] \le \sH_f^K (q) .
$$
We can also see that $\sI_f^K [q;m]$ becomes 0 if message $m$ does not help answer query $q$ as happened in \eqref{sec:ignorance}. 
Note that if $m \in K$, we have $\sI_f^K [q;m] = 0$ (since 
$\sH_f^{K \cup \{m\}} (q) = \sH_f^{K} (q)$ regardless of $q$). Consequently, we need to have an additional assumption that $m$ does not belong to $K$ or $\{m\} \cap K = \emptyset$.
In addition, we say that message $m \notin K$ is independent of query $q$ (with respect to knowledge base $K$) if $\sI_f^K [q;m] = 0$.
Clearly, if message $m$ helps answer query $q$ (to some extent), we expect to see that $\sI_f^K [q;m] > 0$. Thus, the semantic mutual information can be used to quantify the increase of semantic information that message $m$ together with knowledge base $K$ can provide for query $q$. 
Then, assuming that Bob is the legitimate receiver and Eve is the eavesdropper,
the semantic secrecy rate for given query $q$ and message $m
\notin \{ K_{\rm B} \cup K_{\rm E} \}$ can be defined as
\be 
C_{K_{\rm B}; K_{\rm E}} [q;m]= 
\left(\sI_f^{K_{\rm B}} [q;m]
- \sI_f^{K_{\rm E}} [q;m]\right)^+,
\ee 
where $(x)^+ = \max\{x,0\}$.
Let 
\be 
\Delta (q; K_{\rm B}, K_{\rm E}) = 
\sH_f^{K_{\rm B}} (q) - \sH_f^{K_{\rm E}} (q),
\ee 
which is the entropy difference between knowledge bases,
$K_{\rm B}$ and $K_{\rm E}$ for given query $q$. If $\Delta (q; K_{\rm B}, K_{\rm E}) > 0$, we can see that $K_{\rm B}$ has less knowledge than
$K_{\rm E}$ for given query $q$, and vice versa.
Then, we can see that the semantic secrecy rate becomes greater than 0 if
\be 
\Delta (q; K_{\rm B}, K_{\rm E}) >
\Delta (q; K_{\rm B} \cup \{m\}, K_{\rm E} \cup \{m\}).
    \label{EQ:DD}
\ee 
The inequality in \eqref{EQ:DD} implies that  message $m$ can improve Bob's knowledge base more than 
Eve's knowledge base  when answering query $q$.

The notion of semantic secrecy rate can be extended to the case that message $m$ may not be reliably received due to TC errors over noisy physical channels. This generalization can allow the integration of  conventional information security with semantic information security. To this end, we can consider that the original message $m$ is modified due to TC errors and received at Bob and Eve as $m_{\rm B}$ and $m_{\rm E}$, respectively. The semantic mutual information at Bob becomes $\sI_f^{K_{\rm B}} [q;m_{\rm B}]$ and at Eve 
$\sI_f^{K_{\rm E}} [q;m_{\rm E}]$. From them, we could generalize the semantic secrecy rate to encompass different physical TC channel conditions of Bob and Eve.









\section{Issues in Designing SC Systems}    \label{S:SCTC}

In this section, we discuss key issues in designing SC systems with questions and their answers. In particular, we focus on interactions between TC and SC.

\subsection{Question: How Can SC Be Structured?} \label{Sec:structured_SC}

\begin{figure}[h]
\begin{center}
\includegraphics[width=\figwidth]{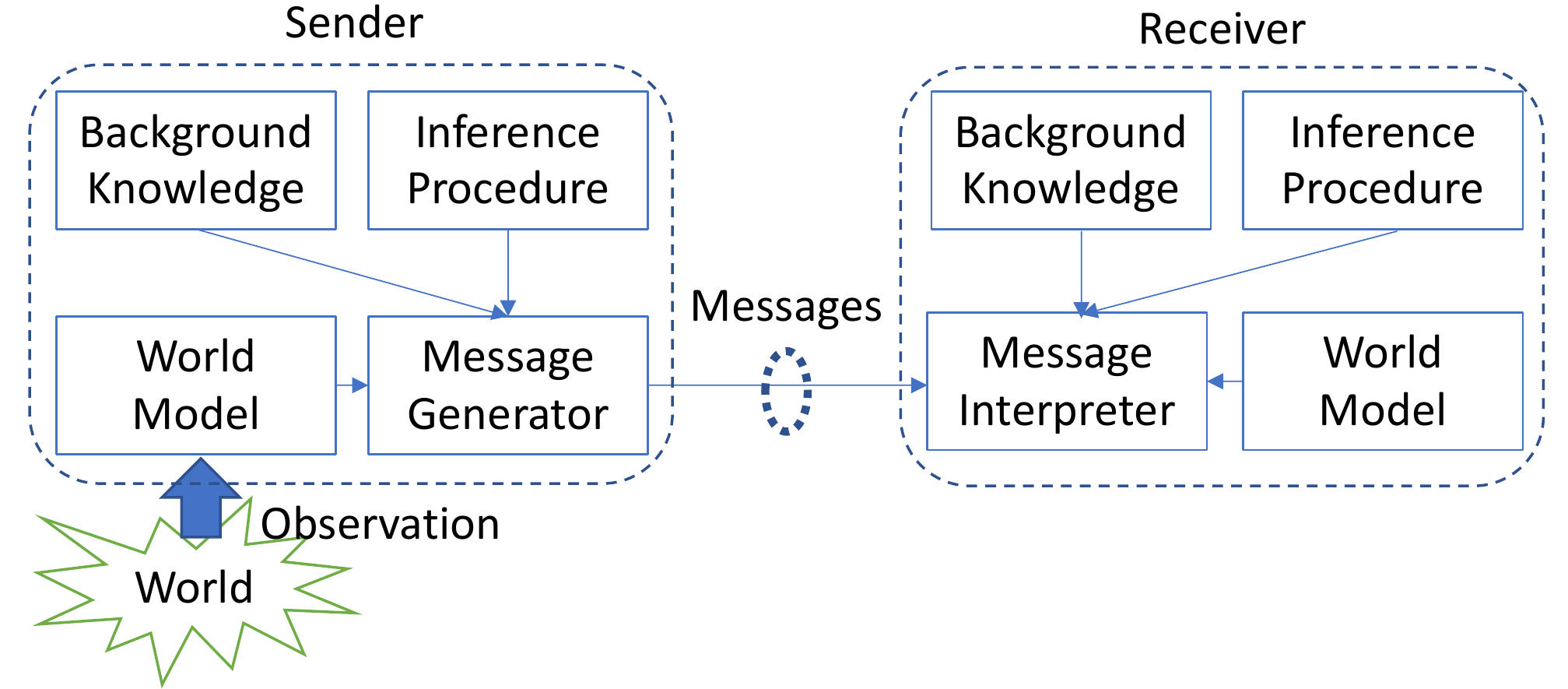} 
\end{center}
\caption{A model of SC from \cite{Bao11}.}
        \label{Fig:Fig1}
\end{figure}

In \cite{Bao11}, a model of SC  was presented, which is illustrated in Fig.~\ref{Fig:Fig1}. The message generator, which is also called a  semantic encoder is to produce a message syntax that will be transmitted by a conventional/technical transmitter. As a result, it is possible to design an SC system with two different layers: 
TC and SC layers.

\begin{figure}[h]
\begin{center}
\includegraphics[width=\figwidth]{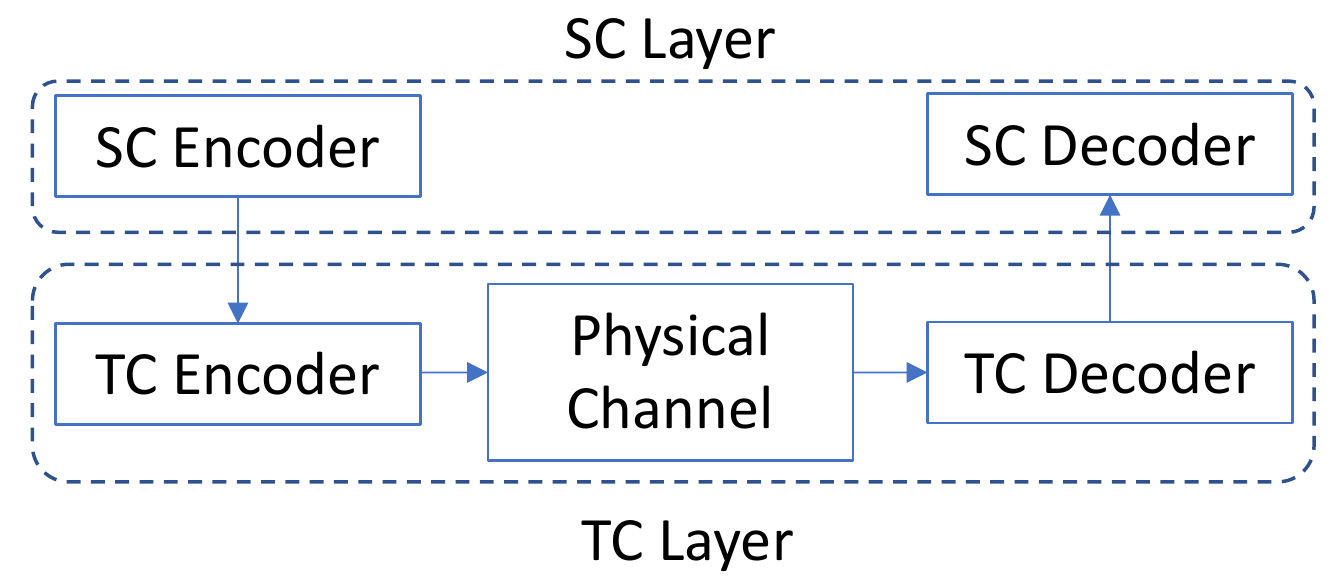} 
\end{center}
\caption{A two-layer model for SC over TC.}
        \label{Fig:Fig2}
\end{figure}

In particular, the output of the sender at the SC layer is a message to be transmitted over a conventional physical channel as shown in Fig.~\ref{Fig:Fig2}. The output of the decoder at the TC layer is a decoded message that becomes the input of the SC decoder. From this view, a conventional TC system can be used without any significant changes for SC. However, without any meaningful interactions between TC and SC, there is no way for TC to exploit the background knowledge in  SC and use the information obtained from semantic inference.

\begin{figure}[h]
\begin{center}
\includegraphics[width=\figwidth]{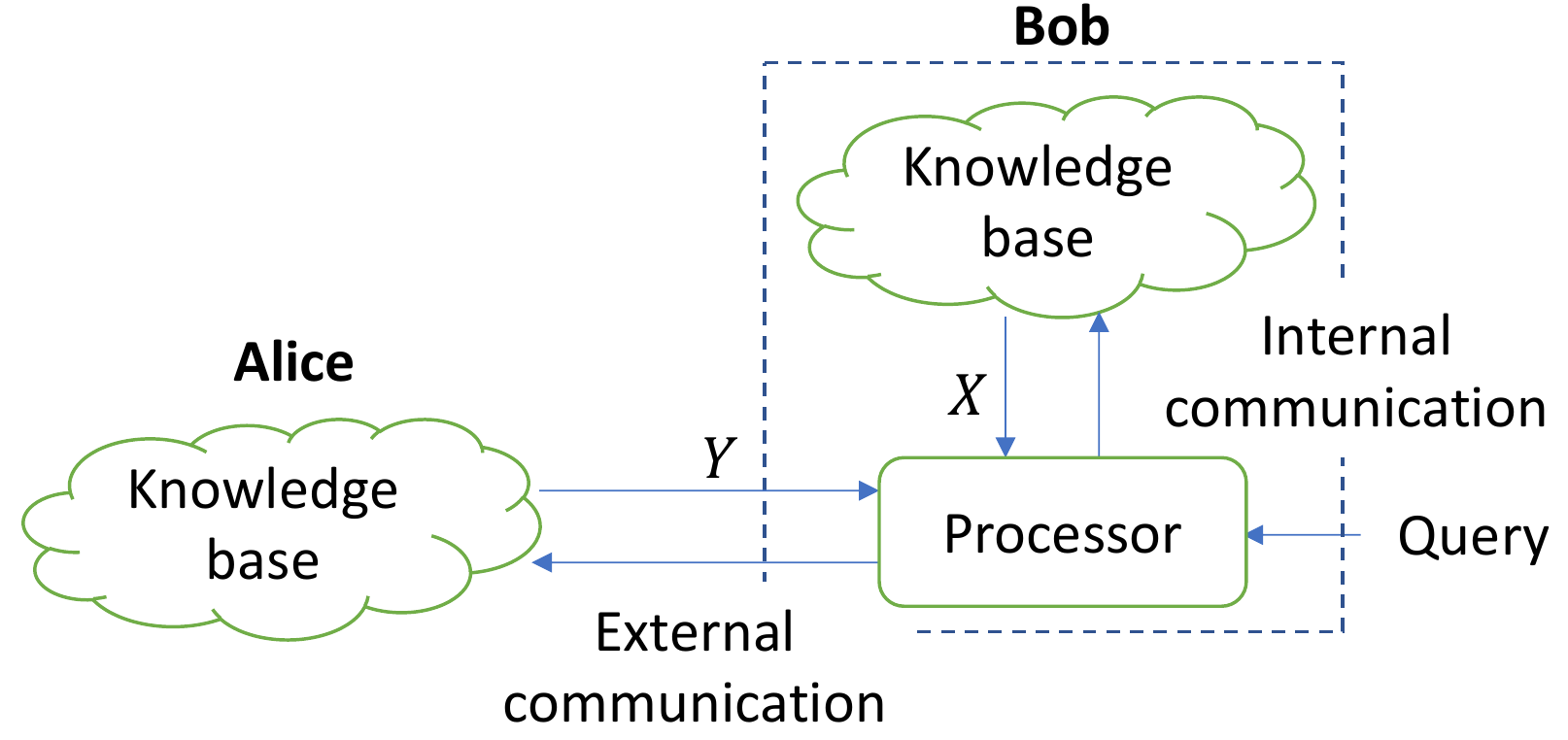} 
\end{center}
\caption{Exploiting the external and internal knowledge bases to reduce the number of bits to transmit.}
        \label{Fig:AB_XY}
\end{figure}

For interactions between TC and SC, the notion of the conditional entropy \cite{CoverBook} can be employed. 
In SC, we can assume that $X$ is the information that can be obtained from the background knowledge at the receiver. 
In particular, $X$ is a clause or an element of clauses in the knowledge base at the receiver. For a clause $X$, the entropy of $X$ becomes $\sH_f (X) = \sH(X)$.
In this case, the sender only needs to send the information of $Y$ at a rate of $\sH(Y\,|\,X)$. In Fig.~\ref{Fig:AB_XY}, we illustrate a model for exploiting the external and internal knowledge bases to reduce the number of bits to transmit. For a given query, Bob can extract partial information, $X$, from his knowledge base, which can be seen as data transmitted through internal communication, and seek additional information, $Y$, from others' knowledge bases, e.g., Alice's knowledge base. In this case, the number of bits to be transmitted is $\sH(Y\,|\, X)$, which will be available through external TC. 

In general,  the notion of the
Slepian-Wolf coding \cite{SW73} can be employed in order to efficiently exploit the background knowledge in SC. 
Suppose that there are two sources at two separate senders, which are denoted by $X$ and $Y$, for distributed source coding. In the Slepian-Wolf coding, sender 1 that has $X$ can transmit $X$ at a rate of $\sH (X)$, while sender 2 that has $Y$ can transmit $Y$ at a rate of $\sH(Y\,|\, X)$, not $\sH(Y)$. 
As a result, the total rate becomes 
$\sH(X) + \sH(Y\,|\, X) = \sH(X,Y) \le \sH(X) + \sH(Y)$. In the context of SC, $X$ can be seen as the information that is available from the background knowledge and through semantic inference. 
\begin{tcolorbox}[enhanced,sharp corners,width={\columnwidth},breakable,colback=white, boxsep=1pt,left=2pt,right=2pt,top=3pt,bottom=3pt]
\begin{myexample}
Suppose that Alice and Bob are the sender
and receiver, respectively. In previous conversations, Alice
told Bob that ``Tom has passed an exam and his score is 75
out of 100," which becomes part of background knowledge. Then, Bob asked Alice the pass score,  which is denoted by $Y$. Clearly, based on the knowledge base from the previous conversation, the pass score has to be less than or equal to 75, i.e., $Y \le 75$, which can be regarded as $X$. Thus, to encode $Y$, the number of bits becomes $\sH(Y\,|\, X)= \sH(Y\,|\, Y \le 75)$. If $Y$ is a positive integer and uniformly distributed over $[1, 100]$, $\sH(Y\,|\, Y \le 75) = 
\sum_{i=1}^{75} \frac{1}{75} \log 75 = \log 75$, not $\sH(Y) = \log 100$.
\end{myexample}
\end{tcolorbox}

\begin{tcolorbox}[enhanced,sharp corners,width={\columnwidth},breakable,colback=white, boxsep=1pt,left=2pt,right=2pt,top=3pt,bottom=3pt]
\begin{myexample}   \label{ex:str2}
Suppose that Eve told Bob that ``Tom's score is 75", which is denoted by fact $a$. In addition, Alice sends additional information that ``The pass score is 70," which is denoted by fact $b$.  Bob still does not know if Tom has passed, even after knowing the mark. Bob can ask Alice but does not need to ask Alice or Eve whether or not Tom has passed, because Bob can tell Tom passes from facts $a$ and $b$ via inference. If $p_a = 0.8$ and $p_b = 0.9$, the probability that Tom has passed is $p_a  p_b = 0.72$. Thus, in order to encode the fact that Tom has passed, which is a binary random variable 
(e.g., $Y = 0$ (resp. $Y = 1$) represents Tom passes 
(resp. fails)), the number of bits becomes 
$\sH_f (a \wedge b) = - (0.72 \log 0.72 + 0.28 \log 0.28) \approx 0.855 < 1$. This demonstrates that the background knowledge in SC can help compress the information in TC. 
A logic programming perspective on this example can also be considered. Suppose we model the knowledge Bob has with this rule that says that a person passes if the mark is above a threshold, and also that Bob has been told by Eve Tom's score:

\begin{verbatim}
0.8::mark(tom,75).
1.0::pass(X) :- 
mark(X,M), pass_score(S), M >=S.
\end{verbatim} 
But Bob still does not know if Tom has passed. Bob could ask Alice but does not need to if he also knows the passing mark:
\begin{verbatim}
0.9::pass_score(70).
0.8::mark(tom,75).
1.0::pass(X) :- 
mark(X,M), pass_score(S), M >=S.
\end{verbatim} 
 Bob can then answer the query {\tt pass(tom)} himself with computed probability $0.72$.
 Now Bob  knows not only Tom's mark but also whether Tom has passed, if this probability of $0.72$ is good enough for Bob. With $K$ representing Bob's knowledge base, note that $\sH^K_f({\tt pass(tom)})=0.593$. 
Note that if Charlie later tells Bob that Tom has passed with probability $0.6$, then Bob perhaps should discard Charlie's message (which under assimilation 
resulting in $K'$) would increase Bob's uncertainty about {\tt pass(tom)} since $\sH^{K'}_f({\tt pass(tom)})=0.673$.
Inferring can go far - e.g., by inferring about Tom, Bob has reduced the need for communication, but this can be extended to not just Tom but many others, saving a lot of communication.  
While this example appears to be contrived, one can consider a wide range of examples where a similar advantage can be realized. For instance,
another way to put it is that  suppose Bob knows the review scores of 1000 restaurants in his city but without knowing the pass score to be qualified as a \emph{good} restaurant. Bob does not know if any of them passed, but on receiving  the one message on the pass score, Bob now can infer which of the 1000 restaurants passed and which did not. 
 Also, rather than sending $1000$ facts stating which passed and which  didn't, sending just the pass score is more efficient. Lastly, if Bob is uncertainty tolerant and guesses the pass score $70$ with probability $0.75$, then it doesn't even need to ask for the pass score, and concludes a restaurant passes with probability $> 0.5$, which might be good enough for tolerant Bob to dine in. 
\end{myexample}
\end{tcolorbox}

\subsection{Question: What Messages to Send?}

As discussed in Subsection~\ref{SS:IRNC}, an optimal message can be chosen to minimize the entropy for a given query $q$  (see \eqref{EQ:mmin}). If a message is to be sent over a TC channel, the length of message can be regarded as the cost of TC. Let $\ell(m)$ denote the length of message $m$, where  $m \in U$ at a sender (in bits),
while $K$ represents the knowledge base at the receiver that has query $q$.
Provided that the maximum length of message is limited by $L_{\rm max}$ over a given TC channel, the optimal message for query $q$ can be given by
\begin{eqnarray}
& m^* = \argmin_{m \in U} \sH_f^{K \cup \{m\}} (q) & \cr 
& \mbox{subject to} \ \ell(m) \le L_{\rm max}. &  
    \label{EQ:mLm}
\end{eqnarray}
While the optimization in \eqref{EQ:mLm} would be tractable, it requires for the sender to know or estimate the receiver's knowledge base, $K$, so that it can compute $\sH_f^{K \cup \{m\}} (q)$. Thus, in general, it is expected that the sender has a larger knowledge base than the receiver and knows (or is able to estimate) the receiver's knowledge base. For example, the sender can be a server in cloud and the receiver can be a mobile user in a cellular system. The server needs to update all the registered users' knowledge bases. In addition, the server is connected to base stations and needs to estimate the length of message $m$ to be transmitted through TC, which may vary depending on the time-varying physical channel condition between the user and associated base station. In this case, $\ell(m)$ is also a function of the channel condition and parameters of the physical layer (e.g., modulation order, code rate, and so on). 

The message selection problem in \eqref{EQ:mLm} can also be generalized for the case of multiple receivers. For example, suppose that there is a common query from all the receivers, $q$.
For TC, we can consider broadcast channels where there one sender and $N$ receivers.
Let $K_n$ denote the knowledge base of receiver $n$, $n \in \{1,\ldots, N\}$. Then, \eqref{EQ:mLm} becomes
\begin{eqnarray}
& m^* = \argmin_{m \in U} \max_n \sH_f^{K_n \cup \{m\}} (q) & \cr 
& \mbox{subject to} \ \ell(m) \le L_{\rm max}, &  
    \label{EQ:gmLm}
\end{eqnarray}
where the maximum of the all receivers' entropy for the common query $q$, i.e., $\max_n \sH_f^{K_n \cup \{m\}} (q)$, is to be minimized. If multiple messages are to be sent, the 
message selection in \eqref{EQ:gmLm} can be repeated or a subset of $U$ can be chosen. 


\subsection{Question: What Questions to Ask?}

As shown in Example~\ref{ex:str2}, it is important to formulate a question/query  carefully in SC for efficient TC. For example, if Bob asks Alice, "Does Tom pass?", Alice can answer yes or no. Thus, a single binary random variable can be considered in TC for Alice's answer. In this regard, inefficient questions might be ``What is the pass score?" and ``What is Tom's score?". Then, Alice must answer "70" and "75", respectively, which requires more than one bit and can be seen inefficient compared to the answer of pass or fail with one binary random variable  in TC. 

In addition, the  knowledge base has to be exploited in SC to formulate a TC-efficient question/query through semantic inference as mentioned earlier. By a TC-efficient question/query in SC, we mean a question/query that can be answered with a minimum number of bits in TC. 
To this end, we can consider the minimum description length (MDL) criterion
\cite{Rissanen78}. 

Recall that
$\mathcal{P}(K)$ denotes the set of queries answerable using knowledge base $K$. Consider a subset of $\mathcal{P}(K)$, denoted by
$\cQ$ (i.e., $\cQ \subseteq \cP(K)$), which has all the queries whose answers can provide specific information that Bob wants.
Then, the query that minimizes the total length of query-and-answer is given by
\be 
q^* = \argmin_{q \in \cQ} \ell(q) + \ell(m\,|\, q),
    \label{EQ:MDL}
\ee 
where $\ell(q)$ and $\ell(m\,|\, q)$ represent the length functions of question $q$ 
and message $m$ (as an answer) for given question $q$, respectively. Furthermore, the cost function can be replaced with $\ell(q) + \lambda \ell(m\,|\, q)$, where $\lambda > 0$ is the weight for the length of answer. In the conventional MDL criterion, $\lambda = 1$. while $\lambda$ can be larger than 1 if the length of answer is more important than the length of question, and vice versa. For a length function, $\ell(\cdot)$, suppose that a set of queries is finite and known, i.e., $\cQ = \{q_1, \ldots, q_N\}$ with a finite $N$ is known. In addition, if the probability that query $q_n$ is given by $\Pr(q_n) = P_n$, using the entropy, then the length of query $q_n$ becomes $L (q_n) = - \log P_n$.
\begin{tcolorbox}[enhanced,sharp corners,width={\columnwidth},breakable,colback=white, boxsep=1pt,left=2pt,right=2pt,top=3pt,bottom=3pt]
\begin{myexample}
Suppose that a group of students took an exam and their scores 
(between 1 and 100) were given.
In addition, there are 4 grades, $\{A,B,C, F\}$. Denoting by $x$ a student's score, the grade is given as follows: $A$ for $x \ge 90$, $B$ for $70 \le x < 90$, $C$ for $50 \le x < 70$, and $F$ for $x < 50$. Alice has a knowledge base of the exam results, and Bob wants to know if Tom has passed and can consider the following set of questions to ask Alice:
\begin{align*} 
\cQ & = \{q_1, q_2, q_3\} \cr
& \!\!\!\!\!\!\!=\{{\tt `Tom's\ score?"}, 
{\tt ``Tom's\ grade?"}, 
{\tt ``Does\ Tom\ pass?"} \}.
\end{align*}
Bob knows the grading table and $F$ means fail. Provided that Tom's score is 80,  the answer that ``Tom's score is 80" (for $q_1$) or ``Tom's grade is B" (for $q_2$) implies that Tom passes. Thus, the answer of any query in $\cQ$ can directly or indirectly provide the information what Bob wants (i.e., whether or not Tom passes).
The number of bits to encode the answer for query $q_1$ is $\log 100 \approx 6.64$ bits (if the score is given as an integer number between 1 and 100 uniformly at random), for query $q_2$ 2 bits (as there are 4 grades that are equally likely), and
for query $q_3$ 1 bit. If $P_1 = 0.6$, $P_2 = 0.3$, and $P_3 = 0.1$, with $\lambda = 1$, then $q^* = q_2$ according to the MLD criterion in \eqref{EQ:MDL}.
Of course, $\ell(m\,|\, q)$ can be shorter than the above values of any prior knowledge of Tom's performance is known (e.g., Tom has been an excellent achiever and hardly fails). This indicates that the receiver's knowledge base can help compress the information to be sent in TC.
\end{myexample}
\end{tcolorbox}

\subsection{Question: How Existing Knowledge Is Related?}

Let $Y$ be a random variable of a certain information.
In addition, denote by $X_k$ the information that user $k$ has. 
Each user may have \emph{a different uncertainty} on the information $Y$ that can be measured by the following conditional entropy:
\be 
W_k = \sH(Y\,|\, X_k) \le \sH(Y).
\ee 
We can decompose the information at user $k$ with respect to $Y$ as follows:
$X_k = (\tilde X_k, \hat X_k)$,
where $\tilde X_k$ is independent of $Y$, i.e.,
$f(Y, \tilde X_k) = f(Y) f(\tilde X_k)$.
Here, $f(X)$ represents the distribution of $X$.
Then, we have
\be 
W_k = \sH(Y\,|\, \tilde X_k, \hat X_k ) =  \sH(Y\,|\, \hat X_k ).
\ee 
Thus, in multiuser SC with single sender and multiple receivers, it becomes important to realize the difference between receivers'  knowledge bases. 
\begin{tcolorbox}[enhanced,sharp corners,width={\columnwidth},breakable,colback=white, boxsep=1pt,left=2pt,right=2pt,top=3pt,bottom=3pt]
\begin{myexample}   \label{ex:3p}
Suppose that $Y$ is a message. The meaning of this message can be different and depends on a receiver's knowledge, which is denoted by $X_k$ for user $k$.
As a result, the meaning of the message is a function of $Y$ and $X_k$, i.e., $S_k = h(Y,X_k)$. Consider 3 parties, Alice, Bob, and Eve. All the three parties know a person named Tom who took an examination. Alice has a message, $Y$, that is ``Tom's score is 75" to deliver Bob and Eve. Bob knows that another candidate whose score is 70 passes, which makes Bob deem Tom passed. Therefore, the meaning of the message is that Tom has passed the examination. On the other hand, Eve knows that a different candidate whose score is 80 passes. Thus, Eve still does not know if Tom has passed.
\end{myexample}
\end{tcolorbox}

In this and previous subsections, we discussed two separate questions, while  the two questions can be considered together. For example, if there are multiple related queries, we may consider an optimal order of queries to minimize the number of bits to be transmitted through interactions between TC and SC. To this end, it is necessary to consider the fact that the receiver can update its knowledge base once the answers of the earlier queries are obtained. Using a certain example, we will discuss this issue in Section~\ref{S:NR} with numerical results.

\subsection{Question: Where To Seek Answer Among Distributed Sources?} \label{SS:DAS}

In this subsection, we first discuss an approach to efficiently select distributed sources in terms of the entropy difference minimization \cite{Choi_WCNC20}. Then, we extend this approach with respect to semantic context. 

Suppose that there are multiple senders and one receiver. Let $X_k$ denote the information that sender $k$ has. The receiver has a query and the answer is a function of the variables at the senders, which  is given by
$Y = \phi(X_1, \ldots, X_N)$,
where $N$ stands for the number of senders. 
For a large $N$, with a limited bandwidth, collecting all information from $N$ distributed senders may take a long time. Furthermore, if the $X_n$'s are correlated, it may not be necessary to collect all variables. For efficient data collection from distributed senders/sources (or sensors), 
the notion of data-aided sensing (DAS) has been considered in \cite{Choi_DAS19} \cite{Choi_DAS20}. If only one sender can be chosen in each round, the following selection criterion is proposed in \cite{Choi_WCNC20}:
\be 
n(i+1) = \argmin_{n \in \cI^c (i)} \sH(\cX^c(i)\,|\, \cX(i)) - \sH(X_n\,|\, \cX(i)),
    \label{EQ:ni}
\ee 
where $\cI (i)$ represents the index set of the senders that send their information up to iteration $i$ and $\cX(i)$ is the set of the variables of the senders corresponding to $\cI(i)$. Here, $\cX^c$ stands for the complement of a set $\cX$. In \eqref{EQ:ni}, $\sH(\cX^c(i)\,|\, \cX(i))$ represents the total amount of remained uncertainty of $\bx = [X_1, \ldots, X_N]$ for given 
$\cX(i)$, which is available at the receiver up to iteration $i$.  Thus, in the next iteration $i+1$, the sender that minimizes the remained uncertainty is to be chosen. 

While
no semantic information is taken into account in \eqref{EQ:ni}, it is possible to extend to consider semantic information. 
Let $m_n$ represent the message at node $n$ (for a given set of queries). At iteration $i$, $K(i)$ represents the updated knowledge base $K$. Then, from \eqref{EQ:S_K}, the node (or source) selection criterion can be given as follows:
\be 
n(i+1) = \argmin_{n \in \cI^c (i)} \left\{ U_{KB} (K(i) \cup \{m_n\} ) - 
U_{KB} (K(i)) \right\}.
    \label{EQ:S_ni}
\ee 
That is, the receiver can actively seek the most effective message among multiple sources and iterate this process to rapidly improve the knowledge base.  In addition, as in \eqref{EQ:mLm}, constraints on TC can be imposed if TC channels are limited (e.g., in terms of capacity and channel resource sharing).


\section{Numerical Results} \label{S:NR}

In this section, we present the numerical results of two examples, illustrating how SC and TC can interact to reduce the communication overhead in terms of the number of bits to transmit or the number of communication rounds. For simplicity, we consider a peer-to-peer communication between Alice and Bob who are the sender and the receiver, respectively, and focus only on the transmission from Alice to Bob. The first example assumes that all the queries from Bob to Alice are assumed to be reliable, whereas the second example considers unreliable queries, as we shall elaborate next.

\subsection{Crossword Puzzle Example} \label{sec:Crossword}
\begin{figure}[h]
\centering
\includegraphics[width=\columnwidth]{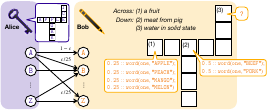} 
\caption{A crossword puzzle scenario.  }
        \label{Fig:cw}
\end{figure}

Consider a task for Bob to solve the crossword puzzle in Fig.~\ref{Fig:cw} with the three questions.
Bob has a knowledge base, denoted by $K_{\rm B}$, to solve the puzzle and is able to ask Alice to obtain answers through TC. It is assumed that Alice knows all the answers, which are (1) APPLE, (2) PORK, and (3) ICE. The physical channel of TC is modeled as a discrete memory less channel (DMC) and TC unit is an alphabet letter (upper cases only). Thus, we assume that each symbol has a unit length of $L_{\rm tc} = 
\log_2 26 \approx 4.7$ bits. For the $26$-ary DMC of TC, the following transition probability is assumed: 
\begin{align*}
y = \left\{
\begin{array}{ll}
x, & \mbox{with a probability of $1- \epsilon$} \cr 
x^\prime \ne x, & \mbox{with a probability of $\frac{\epsilon}{26- 1} =\frac{\epsilon}{25}$}, \cr 
\end{array} 
\right.
\end{align*}
where $\epsilon$ represents the crossover probability or symbol error rate of TC.

To solve the crossword puzzle in Fig.~\ref{Fig:cw},
for problem (1), Bob has a list of possible answers (the names of fruits consisting of 5 letters) as follows, written as an annotated disjunction representing a probability distribution:
\begin{verbatim}
0.25::word(one,"APPLE");
0.25::word(one,"PEACH");
0.25::word(one,"MANGO");
0.25::word(one,"MELON"). 
\end{verbatim}
where ``;'' can be taken as XOR, that is, there are 4 fruits and each one is equally likely in $K_{\rm B}$. For problem (2), Bob has the possible answers as follows:
\begin{verbatim}
0.5::word(two,"BEEF");
0.5::word(two,"PORK").
\end{verbatim}
Bob does not have any idea on problem (3). 
Note that any answer to the query $one(X)$ has probability 0.25 and any answer to the query $two(Y)$ has probability 0.5, that is, $\sH^{K_B}_f(word(one,X)) \approx 0.811$, for each of the possible $X$ and $\sH^{K_B}_f(word(two,Y)) = 1$ for each of the possible $Y$. 
However, we can also capture how knowing certain possibilities in one word can help Bob know the other word, e.g., with certainty 1.0, not labelled below, we have the rules:
\begin{verbatim}
% (1) helps (2)
word(two,"PORK") :-
    word(one,"APPLE").

% (3) helps (1)
word(one,"APPLE") :- 
    word(three,X), endswith(X,"E").
\end{verbatim}


Firstly, we assume that the TC channel is error-free (i.e., $\epsilon = 0$). Without any interactions between TC and SC, Alice needs to send 9 letters or $9 L_{\rm tc}$ bits. To reduce the number of bits from Alice to Bob, Bob can exploit  his knowledge base. To this end, $3! = 6$ orders for queries\footnote{In this example, the terms, problem and query, are interchangeable.} can be considered (e.g., (1) $\Rightarrow$ (2) $\Rightarrow$ (3)). When (1) is asked to Alice, Alice sends ``APPLE". Then, Bob can find the answer of (2) using his knowledge base. Furthermore, Bob can update his knowledge base as ``If the meat is from pig, it is PORK" with probability 1 or 
 
\begin{verbatim}
word(two,"PORK") :- 
    clue(two,"meat from pig").
\end{verbatim}

Bob can ask (3) to Alice and Alice sends the first two letters, ``IC" as the last letter was sent. Note that once (1) is answered, Bob can find the answer of (2). Thus, the two orders of queries, [(1) $\Rightarrow$ (2) $\Rightarrow$ (3)] and  [(1) $\Rightarrow$ (3) $\Rightarrow$ (2)], are reduced to 
[(1) $\Rightarrow$ (3)].
As a result, according to the order of queries, (1) 
$\Rightarrow$ (3), a total of 7 letters should be sent from Alice.
For different orders, we have different number of letters to be transmitted as shown in Table~\ref{TBL:O}.

\begin{table}[thb]   
    \centering
\begin{tabular}{c||c|c}
 & the order of queries & number of letters \cr \hline \hline 
Order 1 & [(1) $\Rightarrow$ (3)] & 7  \cr \hline
Order 2 & [(2) $\Rightarrow$ (3)] & 6  \cr \hline
Order 3 & [(3)] & 3  \cr 
\end{tabular}
    \caption{Total numbers of letters to be transmitted for 3 different orders.} \label{TBL:O}
\end{table}

Note that when problem (3) was asked as the first query, Bob receives ``ICE". Then, he can use his knowledge base to find all the answer. That is, for (1), ``APPLE" is only the answer whose last letter is ``E" so that Bob can find the answer. Likewise, Bob can also find the answer of query (2). This result can provide an insight into the best order for multiple related queries, which is that the  first query might be the most uncertain one for Bob. 
However, this may not be true if TC is no longer error-free (as we will show later).  In addition, note  that the rule that (1) helps (2) could have been more precisely stated that the answer to (1) with the third letter (index 2) is ``P'', then Bob would know word (2):
\begin{verbatim}
word(two,"PORK") :-
    word(one,X), charAt(X,2,"P").
\end{verbatim}
and one could also state that knowing just the third letter of (1) would identify (1) completely:
\begin{verbatim}
word(one,"APPLE") :-
    word(one,X), charAt(X,2,"P").
\end{verbatim}
While this could help further reduce the amount of data Alice needs to send to Bob - Alice just sends the third letter of (1) to identify both (1) and (2), given Alice knows about Bob's knowledge base, we will not consider these more precise rules further for simplicity.

We now consider the case that the crossover probability of TC channel is non-zero (i.e., $\epsilon > 0$).
Suppose that the majority-logic decoding \cite{LinBook} is employed. For query (1), Bob can successfully decode if any 3 letters out of 5 letters, APPLE, are correctly received. For query (2), we assume that 2 letters are to be sufficient for successful decoding (here, we ignore the case that ``POEF" or "BERK" etc). 
On the other hand, for query (3), all the 3 letters should be correctly received, which can happen with a probability of $(1-\epsilon)^3$.

\begin{figure}[h]
\begin{center}
\includegraphics[width=.9\columnwidth]{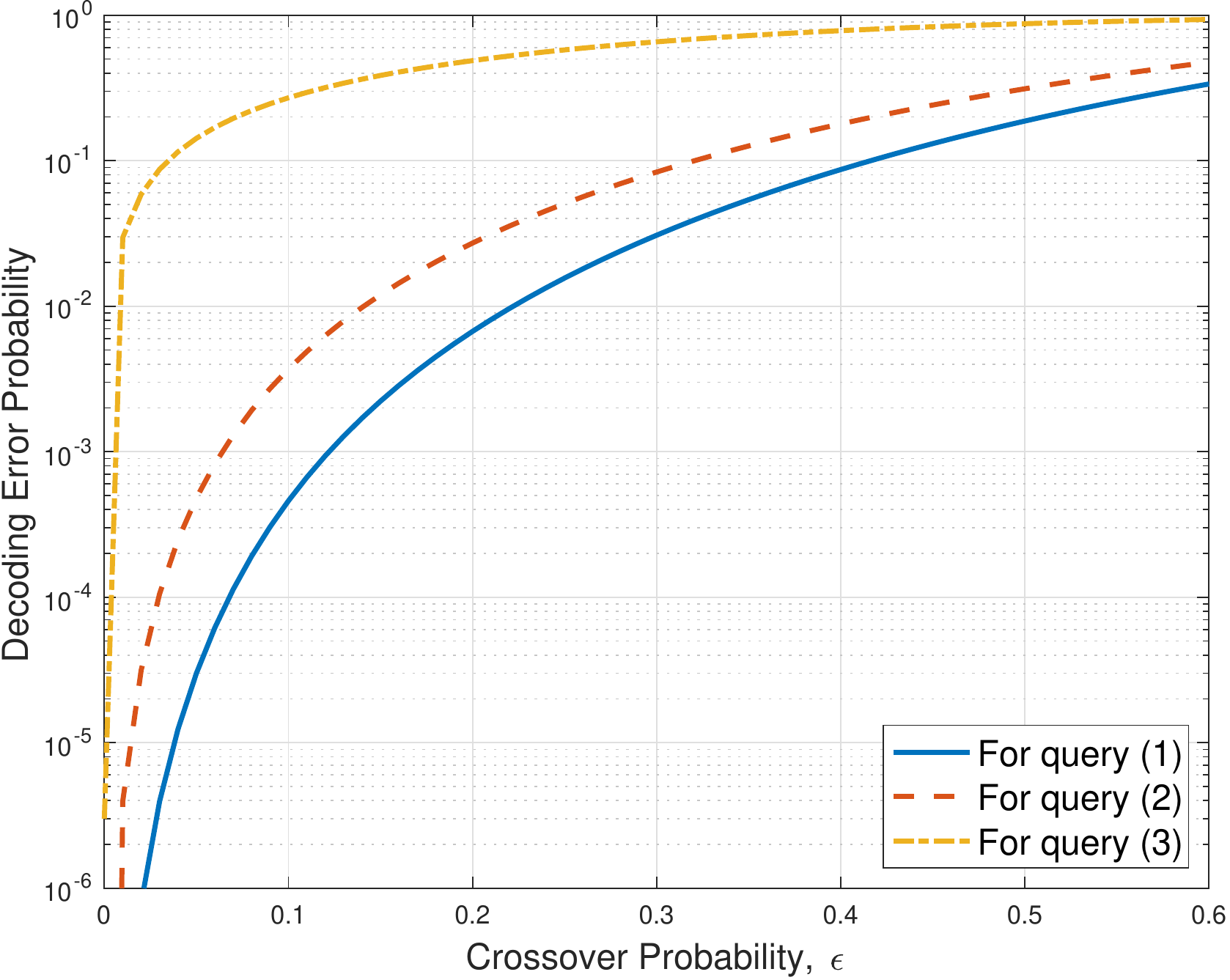} 
\end{center}
\caption{The decoding error probability as a function of crossover probability, $\epsilon$,
when Alice sends the answer for each query.}
        \label{Fig:NR1}
\end{figure}
\begin{figure}[h]
\begin{center}
\includegraphics[width=.9\columnwidth]{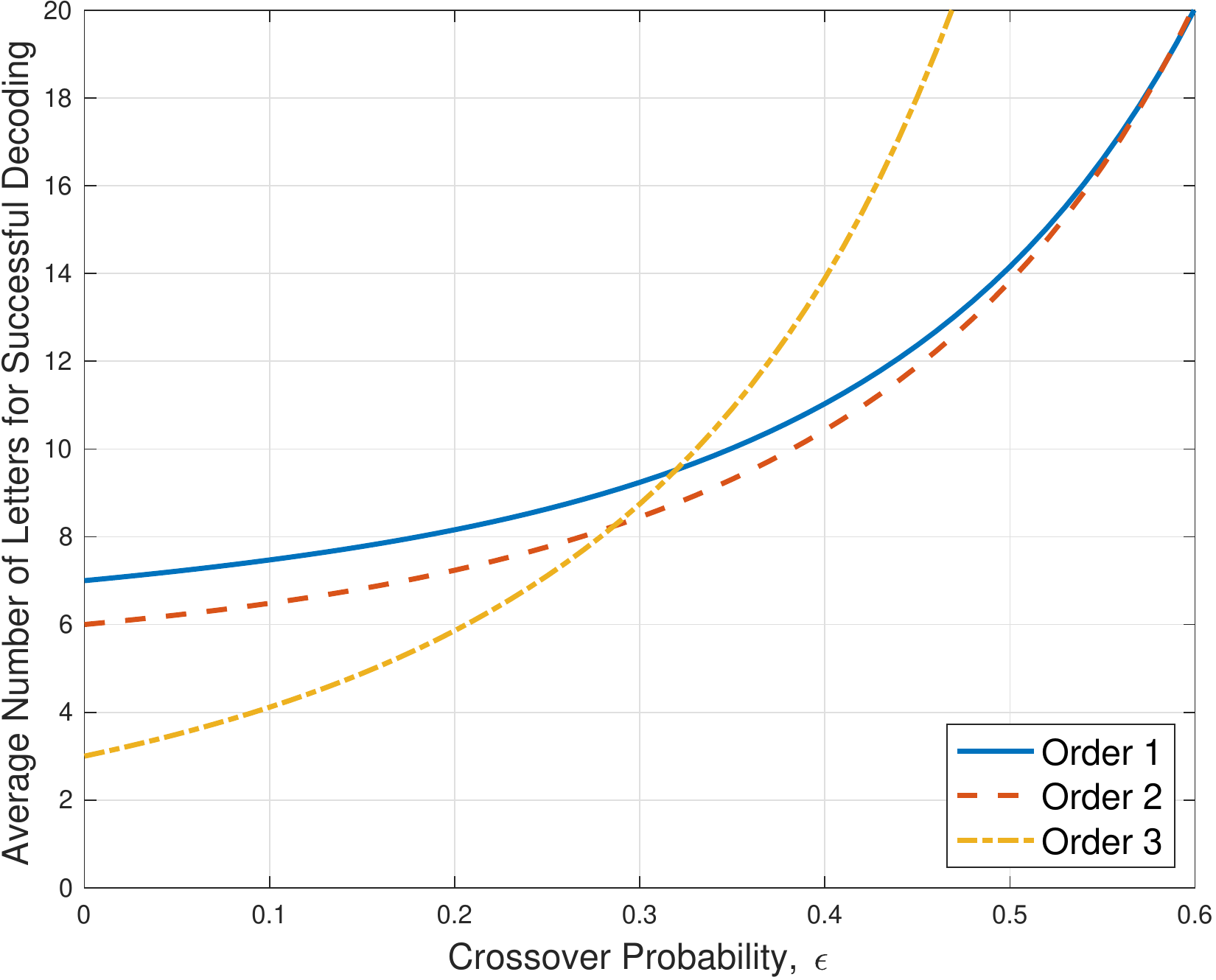} 
\end{center}
\caption{Average number of letters to be transmitted as a function of crossover probability, $\epsilon$, for different orders in queries to solve the crossword puzzle in Fig.~\ref{Fig:cw}.}
        \label{Fig:NR2}
\end{figure}

In Fig.~\ref{Fig:NR1}, the decoding error probability when each query is answered from Alice to Bob over $26$-ary DMC is shown.  Thanks to different levels of knowledge at Bob, the decoding error probability varies. Since Bob does not have any knowledge about query (3), the decoding error probability becomes the highest. When $\epsilon > 0$, there can be decoding errors in TC. Thus, Bob may ask Alice to re-transmit. For this, we consider a simple re-transmission scheme. Then, the average number of re-transmissions can be found using the geometric distribution. For example, letting $p_{\rm s}$  denote the probability of successful decoding for the transmitted answers, the average number of (re-)transmissions becomes $\frac{1}{p_{\rm s}}$, and the total number of letters to be (re-)transmitted becomes $\frac{\mbox{number of letters}}{p_{\rm s}}$ for each query.
Fig.~\ref{Fig:NR2} shows the average number of letters to be (re-)transmitted for each order. It is shown that order 3 is optimal (in terms of the number of letters to be (re-)transmitted) when the crossover probability is sufficiently low (which was clearly shown above when $\epsilon = 0$). However, as the TC channel becomes less reliable
(i.e., $\epsilon$ increases), order 3 is no longer optimal. That is, when $\epsilon \ge 0.29$, we see that order 2 becomes optimal.

\subsection{Clinical Test Example}\label{sec:Clinical}

\begin{figure}
         \includegraphics[width=\columnwidth]{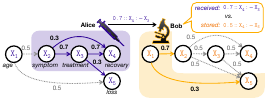}
         \caption{A clinical test scenario.} \label{Fig:clinical}
\end{figure}

Consider a new medicine clinical test in which Alice and Bob participate as a medical doctor and a medical scientist, respectively. As Fig.~\ref{Fig:clinical} shows, Alice in a hospital has a knowledge base storing the causal relationships among a symptom $X_2$, its treatment $X_3$, and a patient's recovery $X_4$; i.e., $p_{23}=p_{34}=0.7$, $p_{24}=p_{35}=0.3$, and $p_{25}=0$, where $p_{ji}:=\Pr(X_i\to X_j)$ and $X_j \to X_i$ is read as `$X_j$ causes $X_i$.' Unknown relationships are associated with random guesses, i.e., $p_{12}=p_{15}=0.5$. Meanwhile, Bob in a lab has a knowledge base storing the causal relationships among the age $X_1$ and loss $X_5$ of the patient; i.e., $p_{12}=0.7$, $p_{15}=0.3$, and $p_{23}=p_{24}=p_{25}=p_{34}=p_{35}=0.5$. 

Such a knowledge base coincides with a causal graph, i.e., a structured causal model (SCM) \cite{scholkopf2021toward} or a Bayesian network, which is a directed acyclic graph (DAG) having the nodes $X_i$'s and the edges associated with $p_{ji}$'s that identify causal relationships. ProbLog is capable of representing this causal knowledge base in a way that ``$X_j \to X_i$ with probability $p_{ji}$" is described by the following clause:
\begin{lstlisting}
$p_{ji}$:: $X_i$:- $X_j$. 
\end{lstlisting}
Given this knowledge base, our focus is Bob's self-asking a query $X_i$ about the truth probability $\Pr(X_i)$ of $X_i$, which is cast as:
\begin{align}
    \Pr(X_i) =1- \prod_{j} \left(1- p_{ji}\Pr(X_j) \right),  \label{Eq:causal}
\end{align}
where $\Pr(X_1)$ is assumed to be $1$. The calculation of $\Pr(X_j)$ follows the same way of \eqref{Eq:causal} in a recursive manner. Consequently, $\Pr(X_i)$ reflects all its preceding causal relationships.

Suppose that answering to each query is followed by improving Bob's knowledge base by receiving a single clause on $p_{ji}$ from Alice. For every communication round, Bob compares the received clause on $p_{ji}$ and the clause $p_{ji}$ stored in its knowledge base. Bob chooses either one of these two clauses and updates its knowledge base. Assuming that the received clause is always chosen by Bob, the communicating clause selection at Alice and the received clause assimilation at Bob are jointly recast as the problem of Alice's selection of a clause to transmit. Each clause transmission is determined by one of the the following rules:
\begin{itemize}
    \item \textbf{A1. Replacement} - A randomly selected clause;
    \item \textbf{A2. Maximum Edge Probability} - The clause associated with the maximal edge probability;
    \item \textbf{A3. Minimum Edge Entropy} - The clause maximally reducing the entropy of an edge in Bob's knowledge base;
    \item \textbf{A4. Minimum Knowledge Base Entropy} - The clause maximally reducing Bob's knowledge base entropy;
    \item \textbf{A5. Maximum Average Answer Probability} - The clause maximizing Bob's average answer probability.
\end{itemize}

With \textbf{A1}, the communication Rounds continue until sending Alice's entire $7$ clauses. With \textbf{A2}-\textbf{A5}, the communication stops when it cannot further improve Bob's knowledge base for its given criterion. This reduces communication costs, which comes at the cost of Alice's additional computing overhead and having the information on Bob's knowledge base.

To measure the accuracy of Bob's reasoning about its query, we define the average error of a query on $X_i$ where the average is taken over the query selection. Each error is measured using the absolute difference between $\Pr(X_i)$ under Bob's knowledge base and that under a ground-truth SCM that can be reconstructed by integrating the knowledge bases of Alice and Bob based on \textbf{A3}. 

\begin{figure}[t]
\centering
 \begin{subfigure}[b]{0.24\textwidth}
         \includegraphics[width=\columnwidth]{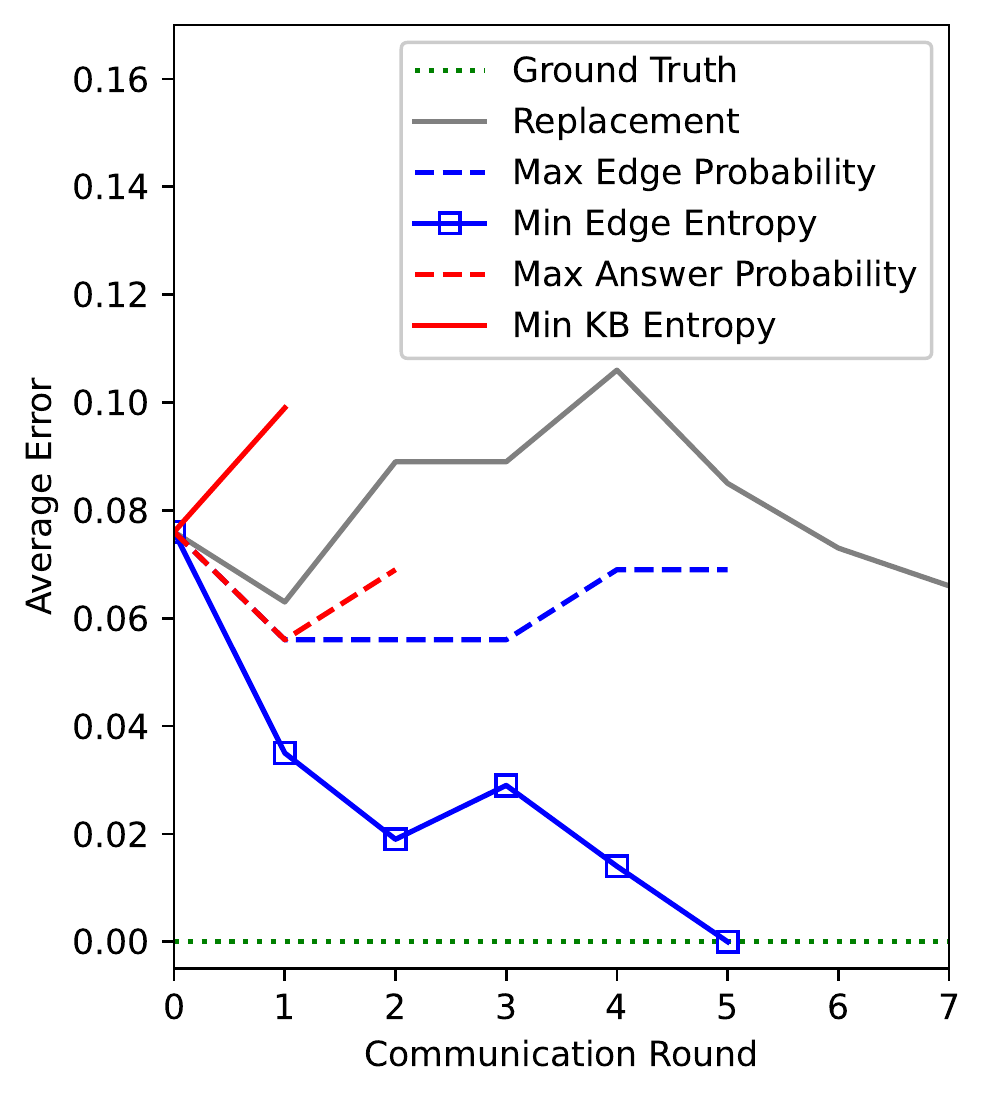} 
         \caption{Average error}
\end{subfigure}
 \begin{subfigure}[b]{0.24\textwidth}
         \includegraphics[width=\columnwidth]{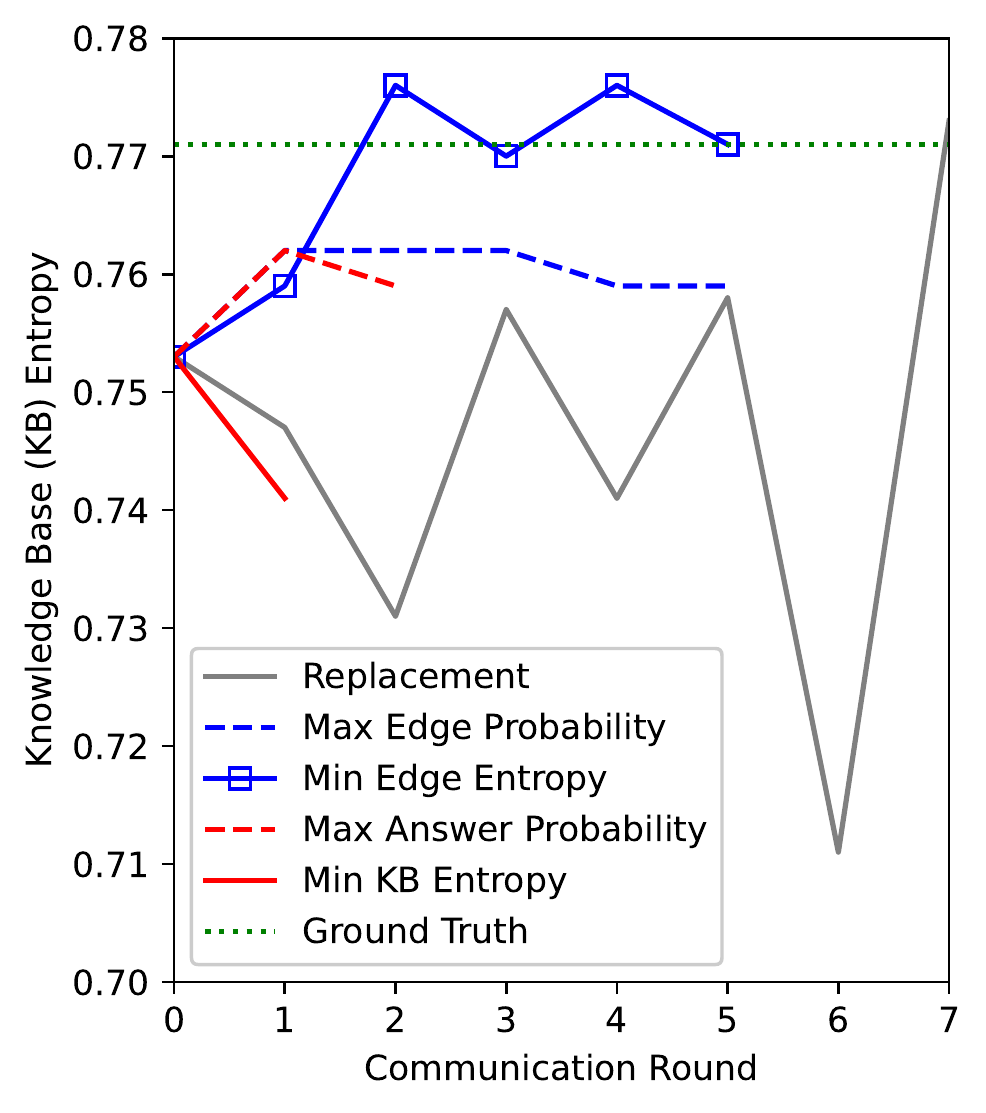}
         \caption{Knowledge base entropy}
\end{subfigure}
\caption{Average error and knowledge base entropy under Alice's transmission rules \textbf{A1}-\textbf{A5}, when Bob's query is randomly selected from $\{X_1,\cdots, X_5\}$. The transmission stops when no further gain is achieved in terms of a given rule.} \label{Fig:Clinical_General}
\end{figure}

When Bob's self-asking queries are randomly selected, Fig.~\ref{Fig:Clinical_General}(a) shows that \textbf{A3} achieves the lowest average error after 5 communication rounds, in stark contrast to \textbf{A2} and \textbf{A5} focusing on the edge/answer probability, corroborating the importance of taking into account entropy. Furthermore, Fig~\ref{Fig:Clinical_General}(b) depicts that \textbf{A3} achieves the entropy of the ground truth SCM, advocating that the knowledge base entropy is a good indicator to identify the reasoning capability of Bob. Nonetheless, the knowledge base entropy is not a proper communication rule for causal reasoning as it ignores the causal relationship therein, as observed by \textbf{A4} that is even worse than \textbf{A1}.

\begin{figure}[t]
\centering
\begin{subfigure}[b]{0.24\textwidth}
         \includegraphics[width=\columnwidth]{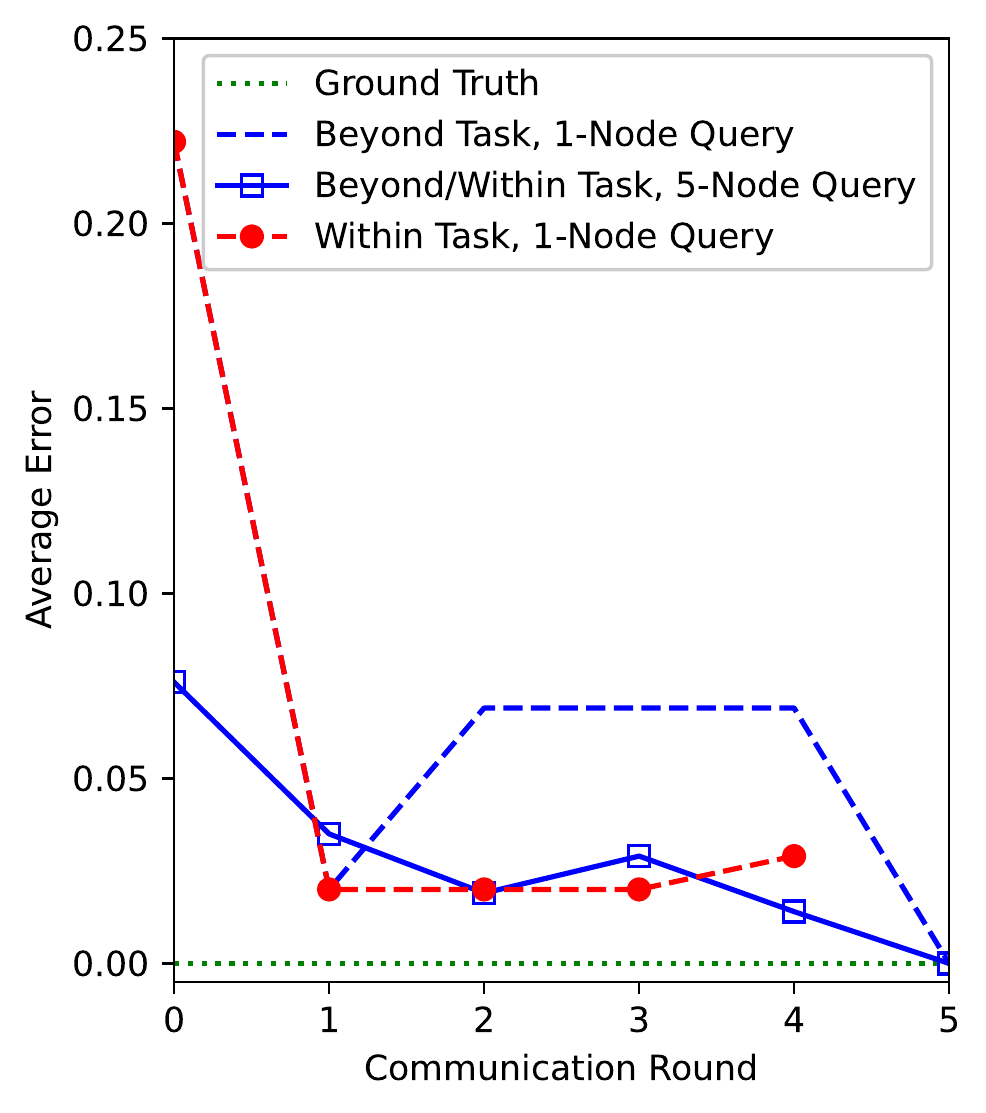}
         \caption{Average Error}
\end{subfigure}
 \begin{subfigure}[b]{0.24\textwidth}
         \centering
         \includegraphics[width=\columnwidth]{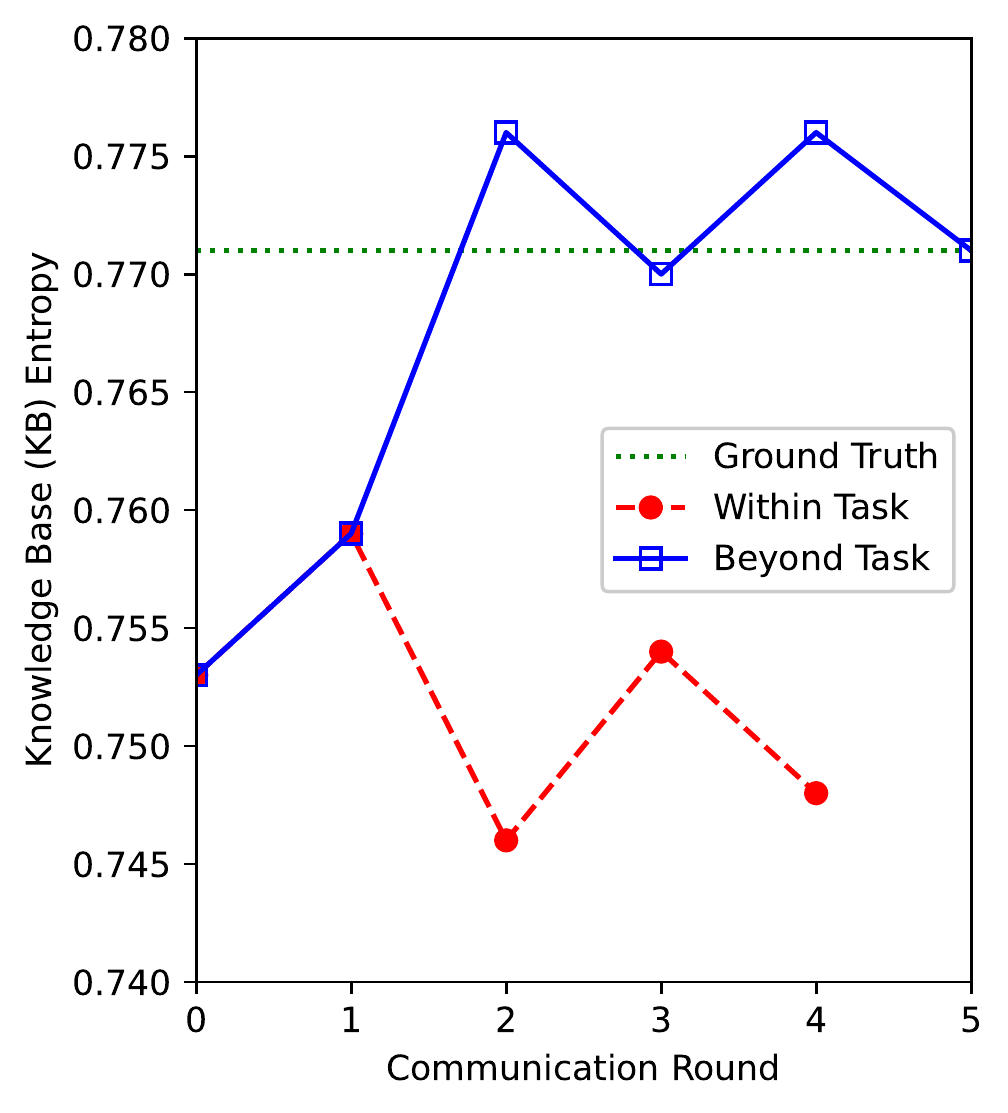}
         \caption{Knowledge base entropy}
\end{subfigure}
\caption{Average error and knowledge base entropy under Alice's transmission rule \textbf{A3} `Beyond Task' and \textbf{A3-1} `Within Task,' when Bob's query is fixed as $X_5$. The transmission stops when  no further gain is achieved in terms of a given rule.}\label{Fig:Clinical_Task}
\end{figure}

Next, we consider that Bob's query is always on $X_5$. Alice can reflect this task specific information in its default transmission rule \textbf{A3} by reducing the target clauses from the its entire knowledge base to only the clauses having $X_5$ as their header. This new rule and the original \textbf{A3} can be interpreted as \textbf{A3-1} `Within Task' and \textbf{A3} `Beyond Task' rules, respectively. Fig.~\ref{Fig:Clinical_Task}(a) shows that \textbf{A3-1} achieves a sufficiently low average error rate with less communication overhead. Nevertheless, as opposed to \textbf{A3}, \textbf{A3-1} fails to achieve the minimum average error due to its ignorance of the causal relationships that are not directly associated with $X_5$. Indeed, the resultant knowledge base entropy under \textbf{A3-1} is different from that under \textbf{A3} and the ground truth value, as observed in Fig.~\ref{Fig:Clinical_Task}(b).

\section{Open Issues and Challenges}    \label{S:OIC}

In this section, we present open issues and challenges to design SC systems.

\subsection{Background Communication for Knowledge Base Updates}

In the previous sections, we have studied SC under a scenario where a user (Bob) has a set of queries to send and ask another user or a server (Alice) who may have a better knowledge base than Bob, to get answers through TC as shown in Fig.~\ref{Fig:AB_XY}. TC may suffer from outages due to fading and interference and from delays due to limited bandwidth. To avoid those difficulties, Bob may update his knowledge base in advance for the \emph{anticipated} queries whenever the bandwidth of TC is sufficient. From this, we can divide TC into: \emph{background TC} for updating the knowledge bases of users and \emph{foreground TC} for sending a query and receiving an answer if the user's knowledge base is not sufficient to obtain the answer with a certain reliability. 

Given limited bandwidth, it is crucial to optimize the resource allocation and scheduling for the foreground TC and background TC. It is expected that the cost of background TC is lower than that of foreground TC as the background TC can be carried out based on best-effort delivery, while the foreground TC needs reliable and low-latency delivery. In this respect, the problem scenario is similar to radio access network (RAN) slicing between ultra-reliable and low-latency communication (URLLC) and other types of services \cite{popovski20185g,bennis2018ultrareliable,pokhrel2020towards}. One key difference is that the priority of background TC depends on the amount of the accumulated knowledge and query patterns and anticipation. In this sense, edge caching problems are also relevant \cite{ko2017live,elbamby2019wireless}. Nonetheless, background TC is additionally challenged by the logical connections within knowledge bases, as observed by the examples in Section~\ref{S:NR}.


\subsection{Pragmatic SC for Memory and Communication Efficiencies}
Thus far we have focused mainly on the Shannon-Weaver's semantics (Level B) problem, while for the effectiveness problem (Level C) we have presumed that all semantic contents can be useful for some generic tasks. Such SC strategies may not be sustainable under limited memory for storing the ever-growing amount of knowledge, not to mention incurring redundant communication costs. Alternatively, inspired from pragmatic information theory \cite{gernert2006pragmatic}, we can first focus on a given task, and then count the usefulness of semantic contents based on its effectiveness in the task. In pragmatic information theory, there is a novelty-confirmation trade-off stating that not only identical information but also too novel information do not contribute to updating knowledge and/or having impacts on decision-makings. While the former is trivial, the latter results from the fact that such dissimilar information is barely comprehensible. 

Leveraging this idea, consider a remote control scenario where Bob updates his knowledge base only when the received clause is grounded in (i) Bob's prior knowledge and (ii) the physical world, directly or through multiple hops. The condition (i) comes from the novelty-confirmation trade-off, and (ii) is based on that control task-effective actions should be taken in the real world. For brevity, consider only conditional clauses in the form of $a\to b$ but an action $u$ that is a factual clause. For given Bob's knowledge base $\{b\to a,  a\to u, u\}$: if Bob receives $c\to b$ satisfying (i) and (ii), Bob updates the knowledge base; and if Bob receives $c\to d$ violating (i) and (ii), Bob keeps the knowledge base unchanged. By adding this pragmatic rule to the aforementioned SC framework, one can communicate and store only the semantic contents that are effective in a given task. In doing so, Bob can save the memory costs by simply discarding less task-effective semantic contents as we studied in the example in Section~\ref{sec:Clinical}. Furthermore, if Alice knows Bob's task effectiveness before transmission, they can save the communication costs too. In this respect, it is worth investigating the feedback and prediction mechanisms to estimate the semantic content's task effectiveness.

\subsection{Compatible SC via Knowledge-Model Conversion}
Our proposed SC layer can be seamlessly added on to the conventional TC layer. How to jointly operate such SC and TC layers have been elaborated in Section~\ref{Sec:structured_SC}, and how to reduce the additional overhead induced by the SC layer will be discussed in Section~\ref{Sec:SC_overhead}. What makes it challenging is the recently proposed semantics-empowered and goal-oriented SC frameworks that commonly rest on AI-native operations with neural networks \cite{qin2021semantic,tong2021federated,liang2022life}, as opposed to our knowledge-based SC layer. We expect that both AI-native and knowledge-based SC frameworks are complementary, even creating a synergetic effect. In this respect, it is promising to study the conversion between neural network models and knowledge bases.

Indeed, it is possible to convert the knowledge base in our SC layer into a neural network model. For instance, treating a knowledge base as a labeled dataset, one can directly infuse the knowlege of the dataset into a neural network model by training the model via supervised learning. Similarly, if the knowledge base is graphical, one can first generate a synthetic corpus from the graph \cite{agarwal2020knowledge}, and train the model, yielding a trained neural network that contains the knowledge of the dataset. On the other hand, it is also feasible to transform a neural network model into a knowledge base, in that the model parameters store the information on their training dataset \cite{achille2019information}. One possible solution is to leverage the model-to-corpus verbalization \cite{west2021symbolic} in natural language processing (NLP), through which a trained model generates synthetic clauses to be stored in a knowledge base. Consequently, an updated knowledge base in our SC layer can improve a neural network model for AI-native SC operations, and vice versa.

\subsection{SC Layer Overhead Reduction via Semantics Alignment} \label{Sec:SC_overhead}
Allocating orthogonal communication and separate computing resources to the SC layer imposes additional overhead on the incumbent communication architecture. A n\"aive solution is superimposing SC and other layers in power domain as in non-orthogonal multiple access \cite{choi2014non,popovski20185g}. Going beyond, one can partly or entirely integrate the SC layer with the existing TC and/or application layers in semantics domain. To illustrate, consider integrating the SC layer message $Z$ into the TC layer message $X$. It requires to maximize $\sI(X;Z) = \sH(X) + \sH(Z) - \sH(X,Z)$. Such a problem boils down to minimizing $\sH(X,Z)$ subject to the fixed marginal distributions of $X$ and $Z$. This coincides with the minimum-entropy coupling problem \cite{kocaoglu2017entropic}, of which the polynomial complexity solution is available \cite{cicalese2019minimum}. Similarly, for the application-SC layer integration, by maximizing $\sI(U;Z) = \sH(U) + \sH(Z) - \sH(U,Z)$, one can align the action $U$ in a control application with $Z$, and vice versa. 

Accordingly, engineering the semantic representation of $Z$ by modifying the logic-based language or learning a new emergent language could be an interesting research direction. As shown by the mutual information expressions above, the SC-TC layer integration and the application-SC layer integration may require more bandwidth due to the increased $\sH(X)$ and incur more uncertain action decision-makings due to higher $\sH(U)$. Furthermore, in different layers, the message sizes can be different, and their communication frequency can be asynchronous. While reflecting this, reducing the SC layer communication overhead via cross-layer integration could be a challenging yet interesting topic for future research.

\subsection{Practical Demonstrators and Practically Establishing Communication Contexts}
We outlined a number of issues with integrating semantics into communication, and noted how  the {\em communication context} (as constituted by the knowledge held by the communicating parties and by what knowledge one party thinks the other has) can help in compression, security and improve efficiency beyond traditional communication models. An analogy is this: one can hear every word (or see every symbol) shared within a conversation among two friends and may not understand what actual knowledge has been exchanged - compression, security (in part at least) and efficiency are concurrently achieved, once the communication context has been established. A practical demonstration of our approach could be useful, e.g., involving  machine-to-machine (say, robot-to-robot, or among IoT devices) communication within a shared context, exploiting such an SC based approach, can help shed further light on the quantitative advantages of our approach. 
One can also investigate how to efficiently establish and maintain such communication contexts before further intensive communications take place.

\subsection{Beyond ProbLog}

We have used ProbLog as a concrete illustration of key ideas of what we mean by semantics and as a way to model semantic information and inference, and to demonstrate how TC and SC can interact information-theoretically. However, there are many types of inferences possible and other logics that can be used to model  semantics. An open research issue is to consider a similar analysis as we have done in this paper but based on a different logical formalism. For example, a generalized version of ProbLog that allows probabilistic argumentation based reasoning ~\cite{DBLP:journals/corr/abs-2110-01990} can help deal with the open world of communication where received messages may support or attack certain other pieces of knowledge, where truth and falsity of statements might not be assumed absolute but weighted by evidence and argument. 

\section{Concluding Remarks}    \label{S:Conc}

While several approaches exist to study semantic information, in this paper, we have considered semantic information and knowledge bases based on probabilistic logic, because the probabilistic logic based approaches can allow us to model interactions between SC and TC and formulate various problems to design a SC system subject to constraints of physical channels in a unified manner. In particular, based on probabilistic logic, we have defined various entropy-based measures for knowledge bases  and addressed various issues when SC and TC layers interact. Numerical examples have been presented to demonstrate how the proposed probabilistic logic based approaches can efficiently utilize TC channels for SC. 

Although we mainly focused on SC between human communicating parties, the proposed approach can be extended to machine-to-machine and human-to-machine SC.
For human communicating parties, in general, we have assumed that one party would improve his/her knowledge (base) by receiving answers to a series of queries in this paper. For machines (in general, autonomous agents), there might be given goals to achieve and SC can be carried out to achieve those goals. Thus, together with the open issues in Section~\ref{S:OIC}, it would be interesting to generalize the proposed approach to SC 
between machines and between machines and human/machine agents.


\bibliographystyle{ieeetr}
\bibliography{si}

\end{document}